\documentstyle[12pt]{article}
\newcommand{\bea}{\begin{eqnarray}}
\newcommand{\eea}{\end{eqnarray}}
\newcommand{\be}{\begin{equation}}
\newcommand{\ee}{\end{equation}}
\newcommand{\vs}[1]{\vspace{#1 mm}}

\newcommand{\dsl}{\pa \kern-0.5em /}

\newcommand{\pa}{\partial}

\newcommand{\nn}{\nonumber\\}

\begin{document}
\topmargin 0pt
\oddsidemargin 0mm



\begin{flushright}

USTC-ICTS-09-05\\



\end{flushright}

\vspace{2mm}

\begin{center}

{\Large \bf The open string pair-production rate enhancement by a
magnetic flux

}

\vs{6}

{\large J. X. Lu\footnote{E-mail: jxlu@ustc.edu.cn}
 and Shan-Shan Xu\footnote{E-mail: xuss@mail.ustc.edu.cn}}

 \vspace{4mm}

{\em

  Interdisciplinary Center for Theoretical Study\\

 University of Science and Technology of China, Hefei, Anhui
 230026, China\\




 }

\end{center}

\vs{15}

\begin{abstract}
We extend the amplitude calculations of \cite{Lu:2009yx} to exhaust
the remaining cases for which one set of D$_p$ branes carrying a
flux (electric or magnetic) is placed parallel at separation to the
other set carrying also a flux but with the two fluxes sharing at
most one common field-strength index. We then find that the basic
structure of amplitudes remains the same when the two fluxes share
at least one common index but it is more general when the two fluxes
share no common index. We discuss various properties of the
amplitudes such as the large separation limit, the onset of various
instabilities and the open string pair production. In particular,
when one flux is electric and weak and the other is magnetic and
fixed, we find that the open string pair production rate is greatly
enhanced by the presence of this magnetic flux when the two fluxes
share no common field-strength index and this rate becomes
significant and may have observational consequences when the
separation is on the order of string scale.

\end{abstract}
\newpage
\section{Introduction}

One type of non-perturbative solitonic objects in superstring
theories (see, for example, \cite{Duff:1994an}) is nowadays called
D-branes \cite{Polchinski:1995mt}. The lowest order stringy
interaction between two such parallel Dp-branes separated by a
distance can be computed either as an open string one-loop annulus
diagram with one end of the open string located at one D-brane and
the other end at the other D-brane or as a closed string tree-level
cylinder diagram with one D-brane, represented by a closed string
boundary state, emitting a closed string, propagating for a certain
amount of time and finally absorbed by the other D-brane, also
represented by a closed string boundary state. When the two D-branes
are at rest, the net interaction vanishes by making use of the usual
``abstruse identity" \cite{Polchinski:1995mt} and this goes by the
name of ``no-force" condition, which usually indicates that the
underlying system preserves certain number of spacetime
supersymmetries.

In addition to the simple strings or simple D-branes, i.e., extended
objects charged under only one NS-NS potential or one R-R potential,
there also exist their supersymmetry preserving bound states such as
(F, D$_p$) \cite{Witten:1995im, Schmidhuber:1996fy, Arfaei:1997hb,
Lu:1999qia, Lu:1999uca, Lu:1999uv, Hashimoto:1997vz,
DiVecchia:1999uf} and (D$_{p - 2}$, D$_p$)
\cite{Breckenridge:1996tt, Costa:1996zd, Di Vecchia:1997pr}, i.e.,
extended objects charged under more than one potential. It would be
interesting to know how to compute the forces between two such bound
states separated by a distance. Since each of the bound states
involves at least two kinds of branes, the force structure is richer
and more interesting to explore. Our focus here will be on the above
mentioned two types of the so-called non-threshold BPS bound states,
namely (F, D$_p$) and (D$_{p - 2}$, D$_p$), with even $p$ in IIA and
odd $p$ in IIB, respectively.

The non-threshold BPS bound state (F, D$_p$), charged under both
NS-NS 2-form potential and R-R $(p + 1)$-form potential, is formed
from the fundamental strings and D$_p$ branes by lowering the system
energy through dissolving the strings in the D$_p$ branes, turning
the strings into a worldvolume electric flux $F_{0a}$ with the flux
pointing along the direction of the original strings. The similar
picture applies to the non-threshold BPS (D$_{p - 2}$, D$_p$) bound
state charged under both R-R $(p - 1)$-form potential and R-R $(p +
1)$-form potential, where the initial D$_{p - 2}$ branes dissolve in
D$_p$ branes, giving rise to a worldvolume magnetic flux $F_{ab}$
with the spatial directions a and b pointing along the codimension-2
directions of the original D$_{p - 2}$ branes inside the D$_p$
branes. Dirac charge quantization implies that the two potentials
for either bound state are characterized by their corresponding
quantized charges, therefore each bound state is characterized by a
pair of integers $(m, n)$. When the pair of integers is co-prime,
the system is stable (otherwise it is marginally unstable)
\cite{Schwarz:1995dk}.

In a previous paper \cite{Lu:2009yx}, the present authors along with
the other two used the description of a boundary state with a
quantized world-volume flux given in \cite{Di
Vecchia:1997pr,DiVecchia:1999uf, Di Vecchia:1999fx} for the bound
state and computed the tree-level cylinder diagram interaction
amplitude between two (F, D$_p$) or between two (D$_{p - 2}$, D$_p$)
bound states when the two bound states are placed parallel at a
separation in a sense that the D$_p$ branes in one bound state are
along the same directions as those in the other bound state and so
are the two fluxes.  In the present paper, we will extend the
computations to exhaust the remaining cases for which the two sets
of D$_p$ branes are still parallel at a separation but the two
fluxes point differently. Concretely we will consider: 1) the two
bound states are both (F, D$_p$) but with their respective
non-vanishing quantized electric fluxes $F_{0a}$ and $F_{0b}$
pointing in a different direction, i.e., with $a \neq b$; 2) the two
bound states are both (D$_{p - 2}$, D$_p$) but with the respective
non-vanishing quantized magnetic fluxes $F_{ab}$ ($a < b$) and
$F_{cd}$ ($c < d$) sharing at most one common index, i.e., either $a
= c$ but $b \neq d$ or $a = d$ or $b = d$ but $a \neq c$ or $a \neq
c$ and $b \neq d$; and 3) one bound state is (F, D$_p$) and the
other (D$_{p - 2}$, D$_p$) with the electric flux $F_{0a}$ pointing
along either or neither of the two spatial indices of the magnetic
flux $F_{cd}$, i.e., either $a = c$ or $a = d$ or $a \neq c, d$.
When the two fluxes share one common index (either temporal or
spatial) in each of the above three cases, we find that all the
amplitudes have the similar structure to the one when the two fluxes
share both of their two indices as given in \cite{Lu:2009yx},
therefore with many features in common. However, when the two fluxes
share no common index, the structure is different and more general,
including the aforementioned one as a special case, therefore having
more rich and interesting features.

Given each bound state characterized by a pair of integers $(m_i,
n_i)$ with $i = 1, 2$, we also find that the non-degenerate (i.e.,
$m_i n_i \neq 0$) force is in general attractive when the two fluxes
are both magnetic or when one flux is magnetic and the other
electric with the two sharing one common index and with the magnetic
flux dominating over the electric flux in effect. However, we are
certain that this force is attractive only at large separation when
the two fluxes are both electric or when one flux is electric and
the other magnetic either with the two
 sharing no common index or with the two sharing one common
index and with the electric flux dominating over the magnetic flux
in effect. When the two fluxes share one common index, the
interaction amplitude can vanish only if there are one electric flux
and one magnetic flux present and the string coupling is completely
determined by the two pairs of the quantized charges with each
characterizing the corresponding bound state. When the two fluxes
share no common index, the amplitude can vanish only if the two
fluxes are both magnetic and have the same magnitude. In either
case, the underlying system preserves only 1/4 of space-time
supersymmetries.

We also study the analytic structures of amplitudes under
consideration and for the case with both fluxes magnetic or with the
magnetic flux dominating over the electric flux in effect when the
two shares one common index, the amplitude is real and diverges when
the brane separation is on the order of string scale, signalling the
onset of tachyonic instability.

For each of the remaining cases, i.e., with one electric flux or at
least one dominant electric flux present, the amplitude has an
imaginary part and this gives rise to a non-vanishing rate for open
string pair production. In particular, when the two fluxes share no
common index, i.e., $p \ge 3$,  the rate of pair production of open
strings is greatly enhanced by the presence of this fixed magnetic
flux even when the electric flux is weak. This rate can be very
significant  even before the onset of tachyonic instability from the
real part of the amplitude when the brane separation is on the order
of string scale. This rate is the largest for $p = 3$ and is at
least one-order of magnitude smaller for $p > 3$. This simple fact
may single out the D3 branes against any other $p > 3$ branes so
long this open string pair production and the subsequent
annihilation of the produced open strings are concerned. For this
and the other related, this significantly enhanced rate may have
observational consequences. Both this rate enhancement and the onset
of tachyonic instability are not seen in a similar context when the
two fluxes share at least one common index.

This paper is organized as follows. In the following section, we
will give a very brief recall of the boundary state with a given
external flux, providing the representation for the non-threshold
(F, D$_p$) or (D$_{p - 2}$, D$_p$) bound state.  In section 3, we
calculate the interaction amplitudes at the closed string tree-level
cylinder diagram for those cases as specified above using the closed
string boundary state approach with each state characterized by an
arbitrary pair of integers $(m_i, n_i)$ (i= 1, 2), and study the
underlying properties. In particular, we discuss the open string
pair production rate enhancement when the electric flux is
orthogonal to the magnetic one which requires $p \ge 3$.  We also
make estimations of the rate for a small electric flux and a fixed
magnetic flux, showing the significance of this rate when the brane
separation is on the order of string scale.  Note that this rate is
the largest for $ p = 3$ and is at least one order of magnitude
smaller for any other $p
> 3$. This simple fact may single out the D3 branes against any other
$p > 3$ branes so long this open string pair production and the
subsequent annihilation of the produced open strings are concerned.
For this and the related, this significantly enhanced rate may have
observational consequences. We summarize the results in section 4.

\section{The boundary state}

We in this section  review very briefly what we need about the
boundary state of D-branes with a constant external field on the
world-volume and set the conventions for this paper. A rather
complete account of this is given in \cite{Di Vecchia:1997pr,
Billo:1998vr, DiVecchia:1999uf, Di Vecchia:1999rh, Di
Vecchia:1999fx}.

In the closed string operator formalism, the supersymmetric BPS
D-branes of type II theories can be described by means of boundary
states $|B\rangle$ \cite{Callan:1986bc,Polchinski:1987tu}. For such
a description, we have two sectors, namely NS-NS and R-R sectors,
respectively. Both in the NS-NS and in R-R sectors, there are two
possible implementations for the boundary conditions of a D-brane
which correspond to two boundary states $|B, \eta\rangle$ with $\eta
= \pm$. However, only the following combinations \be |B\rangle_{\rm
NS} = \frac{1}{2}\left[ |B, +\rangle_{\rm NS} - |B, -\rangle_{\rm
NS} \right],\ee and \be |B\rangle_{\rm R} = \frac{1}{2} \left[ |B,
+\rangle_{\rm R} + |B, -\rangle_{\rm R} \right]\ee are selected by
the GSO projection in the NS-NS and in the R-R sectors,
respectively. The boundary state $|B, \eta\rangle$ is the product of
a matter part and a ghost part \cite{Billo:1998vr} as \be |B,
\eta\rangle = \frac{c_p}{2} |B_{\rm mat}, \eta\rangle |B_{\rm g},
\eta\rangle,\ee where \be |B_{\rm mat}, \eta\rangle = |B_X \rangle
|B_\psi, \eta\rangle, \qquad |B_{\rm g}, \eta\rangle =
|B_{gh}\rangle |B_{sgh}, \eta\rangle .\ee The overall normalization
$c_p$ can be unambiguously fixed from the factorization of
amplitudes of closed strings emitted from a disk
\cite{Frau:1997mq,Di Vecchia:1997pr} and is given by \be c_p =
\sqrt{\pi}\left(2\pi \sqrt{\alpha'}\right)^{3 - p} .\ee  As
discussed in \cite{DiVecchia:1999uf}, the operator structure of the
boundary state does not change even with the presence of an external
flux on the worldvolume and is always of the form \be |B_X\rangle =
{\rm exp}\left[-\sum_{n =1}^\infty \frac{1}{n} \alpha_{- n} \cdot S
\cdot {\tilde \alpha}_{ - n}\right] |B_X\rangle^{(0)},\ee and \be
|B_\psi, \eta\rangle_{\rm NS} = - {\rm i}~ {\rm exp} \left[i \eta
\sum_{m = 1/2}^\infty \psi_{- m} \cdot S \cdot {\tilde \psi}_{- m}
\right] |0\rangle \ee for the NS-NS sector and \be |B_\psi,
\eta\rangle_{\rm R} = - {\rm exp} \left[i \eta \sum_{m = 1}^\infty
\psi_{- m} \cdot S \cdot {\tilde \psi}_{- m} \right] |B,
\eta\rangle_{\rm R}^{(0)} \ee for the R-R sector\footnote{The phases
chosen in (7) and (8) are just for the convenience when we compute
the couplings to various bulk massless modes. }. The matrix $S$ and
the zero-mode contributions $|B_X\rangle^{(0)}$ and $|B,
\eta\rangle_{\rm R}^{(0)}$ encode all information about the overlap
equations that the string coordinates have to satisfy, which in turn
depend on the boundary conditions of the open strings ending on the
D-brane. They can be determined respectively
\cite{Callan:1986bc,DiVecchia:1999uf} as \be S = \left(\left[(\eta -
\hat{F})(\eta + \hat{F})^{-1}\right]_{\alpha\beta},  -
\delta_{ij}\right), \ee  \be |B_X\rangle^{(0)} = \sqrt{- \det
\left(\eta + \hat F\right)} \,\delta^{9 - p} (q^i - y^i) \prod_{\mu
= 0}^9 |k^\mu = 0\rangle, \ee for the bosonic sector, and  \be
|B_\psi, \eta\rangle_{\rm R}^{(0)} = \left(C \Gamma^0 \Gamma^1
\cdots \Gamma^p \frac{1 + {\rm i} \eta \Gamma_{11}}{1 + {\rm i} \eta
} U \right)_{AB} |A\rangle |\tilde B\rangle \ee for the R sector. In
the above, the Greek indices $\alpha, \beta, \cdots$ label the
world-volume directions $0, 1, \cdots, p$ along which the D$_p$
brane extends, while the Latin indices $i, j, \cdots$ label the
directions transverse to the brane, i.e., $p + 1, \cdots, 9$. We
also define $\hat F = 2\pi \alpha' F$ with $F$ the external
worldvolume field. Also in the above, we have denoted by $y^i$ the
positions of the D-brane along the transverse directions, by $C$ the
charge conjugation matrix and by $U$ the following matrix \be U
(\hat F) = \frac{1}{\sqrt{- \det (\eta + \hat F)}} ; {\rm
exp}\left(- \frac{1}{2} {\hat
F}_{\alpha\beta}\Gamma^\alpha\Gamma^\beta\right);\ee where the
symbol $;\quad ;$ means that one has to expand the exponential and
then to anti-symmetrize the indices of the $\Gamma$-matrices.
$|A\rangle |\tilde B\rangle$ stands for the spinor vacuum of the R-R
sector. We would like to point out that the $\eta$ in the above
means either sign $\pm$ or the flat signature matrix $(-1, +1,
\cdots, +1)$ on the world-volume and should not be confused from the
content.

Note that the ghost and super-ghost fields are not affected by the
type of the boundary conditions imposed, therefore the corresponding
part of the boundary state remains the same as the one without the
presence of an external worldvolume field and their explicit
expressions can be found in \cite{Billo:1998vr}. We would like to
point out that the boundary state must be written in the $(-1, -1)$
super-ghost picture in the NS-NS sector, and in the asymmetric $(-
1/2, - 3/2)$ picture in the R-R sector in order to saturate the
super-ghost number anomaly of the disk
\cite{Friedan:1986,Billo:1998vr}.

\section{The interaction amplitude calculations}

We now proceed to calculate the cylinder-diagram amplitude between
any two of the non-threshold BPS (F, D$_p$) and/or (D$_{p - 2}$,
D$_p$) bound states at a separation $Y$ using the boundary state
approach for those cases as specified in the Introduction. In
addition, we will use the results to discuss certain properties of
the underlying systems such as the analytic structure of the
respective amplitudes and calculate the rate of pair production of
open strings in the open string channel for those cases involving at
least one electric-like flux.

The interaction vacuum amplitude can be calculated via
 \be \Gamma
= \langle B (m_1, n_1) | D |B (m_2, n_2) \rangle \ee where the bound
state with a constant
 world-volume field in each sector has been given in section 2 and
 is characterized by a pair of integers $(m_i, n_i)$ with $i = 1,
 2$
and $D$ is the closed string propagator defined as \be D =
\frac{\alpha'}{4 \pi} \int_{|z| \le 1} \frac{d^2 z}{|z|^2} z^{L_0}
 {\bar z}^{{\tilde L}_0}.\ee Here $L_0$ and ${\tilde L}_0$ are
the respective left and right mover total zero-mode Virasoro
generators of matter fields, ghosts and superghosts. For example,
$L_0 = L^X_0 + L_0^\psi + L_0^{gh} + L_0^{sgh}$ where $L_0^X,
L_0^\psi, L_0^{gh}$ and $L_0^{sgh}$ represent contributions from
matter fields $X^\mu$, matter fields $\psi^\mu$, ghosts $b$ and $c$,
and superghosts $\beta$ and $\gamma$, respectively, and their
explicit expressions can be found in any standard discussion of
superstring theories, for example in \cite{Di Vecchia:1999rh},
therefore will not be presented here even though we will need them
in our following calculations. The above total vacuum amplitude has
contributions from both NS-NS and R-R sectors, respectively, and can
be written as $\Gamma = \Gamma_{\rm NS} + \Gamma_{\rm R}$. In
calculating either $\Gamma_{\rm NS}$ or $\Gamma_{\rm R}$, we need to
keep in mind that the boundary state used should be the GSO
projected one as given in Eq. (1) or Eq. (2). For this purpose, we
need to calculate first the following amplitude \be \Gamma (\eta',
\eta) = \langle B^1, \eta'| D |B^2, \eta\rangle \ee in each sector
with $\eta' \eta = \pm$ and $B^i = B (m_i, n_i)$. In doing the
calculations, we can set  $\tilde L_0 = L_0$ in the above propagator
due to the fact that $\tilde L_0 |B\rangle = L_0 |B\rangle$, which
can be used to simplify the calculations. Given the structure of the
boundary state in Eq. (3) and Eq. (4), the amplitude $\Gamma (\eta',
\eta)$ can be factorized as \be \Gamma (\eta', \eta) = \frac{n_1 n_2
c_p^2}{4} \frac{\alpha'}{4 \pi} \int_{|z| \le 1} \frac{d^2 z}{|z|^2}
A^X \, A^{bc}\, A^\psi (\eta', \eta)\, A^{\beta\gamma} (\eta',
\eta),\ee where we have replaced the $c_p$ in the boundary state
given in section 2 by $n c_p$ with $n$ an integer to count the
multiplicity of the D$_p$ branes in the bound state. In the above,
\bea && A^X = \langle B^1_X | |z|^{2 L^X_0} |B^2_X \rangle,\qquad
A^\psi (\eta', \eta) = \langle B^1_\psi, \eta'| |z|^{2 L_0^\psi}
|B^2_\psi, \eta \rangle,\nn && A^{bc} = \langle B^1_{gh} | |z|^{2
L_0^{gh}} | B^2_{gh}\rangle, \qquad A^{\beta\gamma} (\eta', \eta) =
\langle B^1_{sgh}, \eta'| |z|^{2 L_0^{sgh}} |B^2_{sgh}, \eta\rangle.
\eea In order to perform the calculations using the boundary states
given in (6)-(8), (10) and (11),  we need to specify the worldvolume
gauge field and the S-matrix given in (9) for both (F, D$_p$) and
(D$_{p - 2}$, D$_p$) bound states, respectively. Let us denote the
boundary states $\langle B^1, \eta'|$ and $|B^2, \eta\rangle$ in
evaluating the amplitude in (15) as BS1 and BS2, respectively.
Without loss of generality, we can always choose the external flux
$\hat F_1$ associated with BS1 the following way for simplicity.
When this boundary state is the type of (F, D$_p$), we choose $\hat
F_1$ as \be \hat F_1 = \left(\begin{array}{cccccc}
0& -f_1& & &&\\
f_1& 0&  & && \\
 &  &\cdot&&& \\
 &  &     &\cdot&&\\
 &  &     &     &\cdot& \\
 &  &     &     &     &0 \end{array} \right)_{(p + 1) \times (p +
 1)}.\ee The corresponding longitudinal part of the S matrix as given in
(9) is now \be {S_1}_{\alpha\beta} = \left(\begin{array}{ccccccc}
- \frac{1 + f_1^2}{1 - f_1^2} &  \frac{2f_1}{1 - f_1^2} & & &&&\\
- \frac{2 f_1}{1 - f_2^2}& \frac{1 + f_1^2}{1 - f_1^2} &  & && &\\
 &   & 1&&&&\\
 &  &    &\cdot&&& \\
 &  &     & &\cdot&&\\
 &  &     &     &&\cdot& \\
 &  &     &     &     &&1 \end{array} \right)_{(p + 1) \times (p +
 1)}.\ee
While for the boundary state being (D$_{p - 2}$, D$_p$), we choose
the $\hat F_1$ as \be \hat F_1 = \left(\begin{array}{cccccc}
0& & & &&\\
& \cdot&  & && \\
 &  &\cdot&&& \\
 &  &     &\cdot&&\\
 &  &     &     &0& -f_1\\
 &  &     &     &  f_1   &0 \end{array} \right)_{(p + 1) \times (p +
 1)},\ee with now the quantized $f_1 = - m_1/n_1$. We then have the longitudinal part of
the S matrix as \be {S_1}_{\alpha\beta} =
\left(\begin{array}{ccccccc}
- 1 &  & & &&&\\
& 1 &  & && &\\
 &   & \cdot&&&&\\
 &  &    &\cdot&&& \\
 &  &     & &\cdot&&\\
 &  &     &     &&\frac{1 - f_1^2}{1 + f_1^2}&\frac{2 f_1}{1 + f_1^2} \\
 &  &     &     &     &- \frac{2 f_1}{1 + f_1^2}&\frac{1 - f_1^2}{1 + f_1^2} \end{array} \right)_{(p + 1) \times (p +
 1)}.\ee

With the above choice for $\hat F_1$, the external worldvolume flux
$\hat F_2$ for BPS2 shall be the following for those cases
considered in this paper. When this boundary state is the type of
(F, D$_p$), the only non-vanishing components are $(\hat F_2)_{0a} =
- (\hat F_2)_{a0} = - f_2$ with the given spatial worldvolume index
$a \neq 1$ when BS1 is also the same type but without such a
restriction on this index when BPS1 is the type of (D$_{p - 2}$,
D$_p$). While for this boundary state being the type of (D$_{p -
2}$, D$_p$), the only non-vanishing components are $(\hat F_2)_{bc}
= - (\hat F_2)_{cb} = - f_2$ for given worldvolume spatial indices
$c$ and $b$ ( $ c
> b$) and with the only restriction $b \neq p - 1$ when BS1 is of the same type
but without any restriction when BS1 is the type of (F, D$_p$). Here
each flux $f_i$ with $i = 1, 2$ is quantized with a pair of integers
($m_i, n_i$) as discussed in \cite{Lu:2009yx}, and is given for the
case of electric flux as \be f_i = - \frac{m_i}
{\triangle^{1/2}_{e(m_i, n_i)}},\ee where \be \triangle_{e(m_i,
n_i)} \equiv m_i^2 +
 \frac{n_i^2}{g_s^2}\ee with $(m_i, n_i)$ a pair of integers,  $g_s$ the
 string coupling constant and the subscript `e' representing the flux being electric, while for the case of
magnetic flux \be f_i = - \frac{m_i}{n_i},\ee and for latter purpose
we also define \be \triangle_{m (m_i, n_i)} = m_i^2 + n_i^2,\ee with
the subscript `m' representing the flux being  a magnetic one.

With the above preparations, we are now ready to perform rather
straightforward calculations for the various matrix elements
specified in (17) in either NS-NS or R-R sector for those cases
under consideration, using (6)-(8), (10) and (11) for the boundary
states with $\hat F$ and the matrix $S$ as just described as well as
the full expression of $S$ as given in (9). The
calculations\footnote{For  $p = 3$, the interaction with only one
boundary state carrying a particular form of flux was considered for
a completely different purpose in \cite{DiVecchia:2003ae}. An
implicit expression for the interaction in general was given in
\cite{Arfaei:1999kr} but the discussion on the R-R sector there is
confusing and we don't agree in this part, in particular on the
regularization procedure of zero-modes.}  can be clarified as two
types according to whether the two worldvolume fluxes $(\hat
F_1)_{\alpha\beta}$ and $(\hat F_2)_{\gamma\delta}$ discussed above
have one common index (either temporal or spatial) or have no common
index at all for which we will discuss each separately next.

\subsection{Two fluxes with one common index}

The common index for the two fluxes just mentioned can be along
either a temporal or a spatial direction and for the present case we
need the worldvolume spatial dimensions $p \ge 2$. For these cases,
the corresponding amplitude has the same structure as the one
obtained in \cite{Lu:2009yx} when the two fluxes share the same
indices. For simplicity, let us denote the electric flux as `e' and
magnetic flux as `m' and so all possibilities in the present case
can be denoted as\footnote{Note that the (m, e) case can be obtained
from the (e, m) case by sending $f_1 \rightarrow i f_1, f_2
\rightarrow i f_2$ in what follows. So for simplicity, we will not
list this case explicitly. Note also that the (m, m) case can be
similarly obtained from (e, e) case with the same replacements.
However, we would like to discuss these two later cases explicitly
since the force nature and other properties of these two cases are
rather different.} (e, e), (e, m), (m, e) and (m, m) which represent
that BS1 and BS2 are both of the type (F, D$_p$), BS1 the type of
(F, D$_p$) while BS2 the type of (D$_{p - 2}$, D$_p$), BS1 the type
of (D$_{p - 2}$, D$_p$) and BS2 the type of (F, D$_p$), and BS1 and
BS2 both of the type (D$_{p - 2}$, D$_p$), respectively.
 We have now the various matrix elements specified in
(17) as
 \bea &&A^X = C_F \, V_{p + 1}\, e^{- \frac{Y^2}{2\pi
\alpha' t}} \left(2\pi^2 \alpha'\, t\right)^{- \frac{9 - p}{2}} \,
\prod_{n = 1}^\infty \frac{1}{(1 - \lambda |z|^{2n})(1 -
\lambda^{-1} |z|^{2n})(1 - |z|^{2n})^8},\nn && A^{bc} = |z|^{-2}
\prod_{n = 1}^\infty (1 - |z|^{2n})^2,\eea for both NS-NS and R-R
sectors, \bea && A_{\rm NS}^{\beta\gamma} (\eta', \eta) = |z|
\prod_{n = 1}^{\infty} \frac{1}{(1 + \eta' \eta\, |z|^{2n -
1})^2},\nn && A^\psi_{\rm NS} = \prod_{n = 1}^\infty (1 + \eta' \eta
\,\lambda |z|^{2n - 1}) (1 + \eta' \eta\,\lambda^{-1}\, |z|^{2n -
1}) (1 + \eta' \eta \,|z|^{2n - 1})^8, \eea for NS-NS sector, and
\be A^{\beta\gamma}_{\rm R} (\eta', \eta) A^\psi_{\rm R} (\eta',
\eta) = - 2^4 \,|z|^2\, D_F\, \delta_{\eta'\eta, + } \prod_{n =
1}^\infty (1 + \lambda\, |z|^{2n}) (1 + \lambda^{- 1} \, |z|^{2n} )
( 1 + |z|^{2n})^6, \ee for the R-R sector. Note that we have $|z| =
e^{- \pi t}$ above, the matrix elements for ghosts and superghosts
are independent of the external fluxes as expected, and in (28) we
have followed the prescription given in \cite{Billo:1998vr,Di
Vecchia:1999rh} not to separate the contributions from matter fields
$\psi^\mu$ and superghosts in the R-R sector in order to avoid the
complication due to the respective zero modes.  Also in the above,
we have \be D^{-1}_F  = C_F = \left\{\begin{array} {cc} \sqrt{(1 -
f_1^2)(1 - f_2^2)} & {\rm for}
\,(e, e), \\
&\\
\sqrt{(1 - f_1^2)(1 + f_2^2)} & {\rm for}
\,(e, m), \\
&\\

  \sqrt{(1 + f_1^2)(1 + f_2^2)} & {\rm for} \,(m, m),\\
\end{array}\right. \ee

and \be \lambda + \lambda^{- 1} =  2 (2 D_F^2 - 1) =
\left\{\begin{array} {cc} 2 \frac{1 + f_1^2  + f_2^2 -  f^2_1
f^2_2}{(1 - f_1^2)(1 - f_2^2)} & {\rm for}
\,(e, e), \\
&\\
2 \frac{1 + f_1^2  - f_2^2 + f^2_1 f^2_2}{(1 - f_1^2)(1 + f_2^2)} &
{\rm for}
\,(e, m), \\
&\\
2 \frac{1 - f_1^2 - f_2^2 - f^2_1 f^2_2}{(1 + f_1^2)(1 + f_2^2)} &
{\rm for} \,(m, m).
\end{array}\right. \ee Note that in both equations above, the other
cases list above can be obtained, for example, from the (e, e) case
simply by sending $f_i$ to its imaginary value if the corresponding
flux is a magnetic one. For (e, e), $D_F > 1$ and for (m, m), $D_F <
1$. For (e, m) or (m, e), when $D_F > 1$ we say that the electric
flux is dominant in effect while $D_F < 1$ we say that the magnetic
flux is dominant in effect.

As discussed in \cite{Lu:2009yx}, in calculating $A^X$ and $A^\psi
(\eta', \eta)$ as given explicitly above, we have made use of an
important property for the S matrix \be {S^T}_\mu\,^\rho
S_\rho\,^\nu = \delta_\mu\,^\nu,\ee with $T$ denoting the transpose.
This property enables us to perform unitary transformations of the
respective operators in the boundary states (6)-(8) such that the
$S$ matrix appearing, for example, in BS1 completely disappears,
while leaving BS2 with a new S matrix as $S = S_2 S^T_1$, in the
course of evaluating the respective $A^X$ or $A^\psi$. This new S
matrix shares the same property (31) as the original $S_1$ and $S_2$
do but its determinant is always equal to one. Therefore this S
matrix under consideration can always be diagonalized to give two
eigenvalues $\lambda$ and $\lambda^{-1}$ with their sum as given in
(30) above and the other eight eigenvalues all equal to one. This is
the basis for the structure appearing in the contributions due to
the respective oscillators to the $A^X$ and $A^\psi (\eta', \eta)$
as given in (26)-(28) above.

We can now have the vacuum amplitude in the NS-NS sector as \bea
\Gamma_{\rm NS}& =& \,_{\rm NS}\langle B^1| D |B^2 \rangle_{\rm NS}
\nn &=& \frac{n_1 n_2 \, V_{p + 1}\, C_F}{2  (8 \pi^2
\alpha')^{\frac{1 + p}{2}}} \int_0^\infty \frac{d t}{t} \, e^{-
\frac{Y^2}{2 \pi \alpha' t}} \, t^{ - \frac{7 - p}{2}} \nn
&\,&\,\qquad\quad \times |z|^{ - 1} \left[\prod_{n = 1}^\infty
\frac{(1 + \lambda\, |z|^{2n - 1})(1 + \lambda^{- 1} \,|z|^{2n - 1})
(1 + |z|^{2n - 1})^6}{(1 - \lambda\, |z|^{2n})(1 - \lambda^{- 1} \,
|z|^{2n}) (1 - |z|^{2n})^6}\right.\nn &\,&\,\qquad\qquad\qquad\left.
- \prod_{n = 1}^\infty \frac{(1 - \lambda\, |z|^{2n - 1})(1
-\lambda^{- 1} \,|z|^{2n - 1}) (1 - |z|^{2n - 1})^6}{(1 - \lambda\,
|z|^{2n})(1 - \lambda^{- 1} \, |z|^{2n}) (1 -
|z|^{2n})^6}\right],\eea where we have used the GSO projected
boundary state in (1) for $|B^i\rangle_{\rm NS}$ (i = 1, 2) with
$B^i$ as defined previously and have made use of the matrix elements
in (26) and (27) and the amplitude in (16). Also we have used in the
above \be \int_{|z| \le 1} \frac{d^2 z}{|z|^2} = 2\pi^2
\int_0^\infty d t, \ee with $|z| = e^{- \pi t}$. The corresponding
vacuum amplitude in the R-R sector is now \bea \Gamma_{R} &=&
\,_{\rm R}\langle B^1| D |B^2 \rangle_{\rm R} \nn &=& - \frac{ 8\,
n_1 n_2\, V_{p + 1}}{ (8 \pi^2 \alpha')^{\frac{1 +  p}{2}}}
\int_0^\infty \frac{d t}{t}\,e^{- \frac{Y^2}{2 \pi \alpha' t}} \,
t^{ - \frac{7 - p}{2}} \nn &\,& \qquad \quad \times \prod_{n =
1}^\infty \frac{(1 + \lambda\, |z|^{2n})(1 + \lambda^{- 1}
\,|z|^{2n}) (1 + |z|^{2n})^6}{(1 - \lambda\, |z|^{2n})(1 -
\lambda^{- 1} \, |z|^{2n}) (1 - |z|^{2n})^6},\eea where we have used
the GSO projected boundary state in (2) for $|B^i\rangle_{\rm R}$ (i
= 1, 2) again with $B^i$ as defined previously  and made use of the
matrix elements in (26) and (28) and the amplitude in (16) as well
as the equation (33). For both of (32) and (34), we have also made
use of \be \frac{c_p^2}{32 \pi (2\pi^2 \alpha')^{\frac{7 - p}{2}}} =
\frac{1}{(8 \pi^2 \alpha')^{\frac{p + 1}{2}}}\times \frac{1}{2},\ee
where we have used the explicit expression (5) for $c_p$. We also
always assume both $n_1 $ and $n_2 $ are positive integers and the
p-branes in the non-threshold bound states are both D$_p$ branes (or
both anti D$_p$ branes). In the case when the p-branes in either of
the non-threshold bound states (but not both) are anti D$_p$ branes,
the corresponding $\Gamma_{\rm R}$ will switch sign from the one
above but the $\Gamma_{\rm NS}$ will remain the same. In what
follows, we will focus on that the p-branes in both non-threshold
bound states are D$_p$-branes, i.e., (34) is valid. The case when
the p-branes in either of the bound states are anti D$_p$-branes can
be similarly analyzed.

The total vacuum amplitude is now \bea \Gamma &=& \Gamma_{\rm NS} +
\Gamma_{\rm R} \nn &=& \frac{n_1 n_2 \, V_{p + 1}\, C_F}{2  (8 \pi^2
\alpha')^{\frac{1 + p}{2}}} \int_0^\infty \frac{d t}{t} \, e^{-
\frac{Y^2}{2 \pi \alpha' t}} \, t^{ - \frac{7 - p}{2}} \nn
&\,&\,\quad \times \left\{ |z|^{ - 1} \left[\prod_{n = 1}^\infty
\frac{(1 + \lambda\, |z|^{2n - 1})(1 + \lambda^{- 1} \,|z|^{2n - 1})
(1 + |z|^{2n - 1})^6}{(1 - \lambda\, |z|^{2n})(1 - \lambda^{- 1} \,
|z|^{2n}) (1 - |z|^{2n})^6}\right.\right.\nn
&\,&\,\qquad\qquad\qquad\left. - \prod_{n = 1}^\infty \frac{(1 -
\lambda\, |z|^{2n - 1})(1 -\lambda^{- 1} \,|z|^{2n - 1}) (1 -
|z|^{2n - 1})^6}{(1 - \lambda\, |z|^{2n})(1 - \lambda^{- 1} \,
|z|^{2n}) (1 - |z|^{2n})^6}\right]\nn &\,&\,\qquad\quad \left. -
2^4\, D_F\,\prod_{n = 1}^\infty \frac{(1 + \lambda\, |z|^{2n})(1 +
\lambda^{- 1} \,|z|^{2n}) (1 + |z|^{2n})^6}{(1 - \lambda\,
|z|^{2n})(1 - \lambda^{- 1} \, |z|^{2n}) (1 -
|z|^{2n})^6}\right\},\eea  which looks in form exactly the same as
the one for the case when BS1 and BS2 are both of the same type,
i.e., either (F, D$_p$) or (D$_{p - 2}$, D$_p$), and the
corresponding two worldvolume fluxes are along the same directions,
as calculated in \cite{Lu:2009yx}. This is part of the basic result
of this paper. This amplitude can also be expressed nicely in terms
of $\theta$-functions and the Dedekind $\eta$-function with their
standard definitions as given, for example, in \cite{polbookone}. We
then have \bea \Gamma &=& \frac{n_1 n_2 \, V_{p + 1} \, C_F\, \sin
\pi \nu}{ (8 \pi^2 \alpha')^{\frac{1 + p}{2}}} \int_0^\infty \frac{d
t}{t} \, e^{- \frac{Y^2}{2 \pi \alpha' t}} \, t^{ - \frac{7 - p}{2}}
\nn &\,&\quad\times \frac{1}{\eta^9 (it)} \left[\frac{\theta_3
(\nu|it)\, \theta^3_3 (0|it)}{\theta_1 (\nu |it)} - \frac{\theta_4
(\nu | it) \theta^3_4 (0 | it)} {\theta_1 (\nu | it)} -
\frac{\theta_2 (\nu | it) \theta^3_2 (0 | it)}{\theta_1 (\nu | it)}
\right],\eea where we have defined $\lambda = e^{2\pi i \nu}$ and
used the fact $\cos\pi \nu = D_F = C^{-1}_F$ which can be obtained
from $\lambda + \lambda^{-1} = 2 (2 D_F^2 - 1)$ as given in (30)
with $C_F$ and $D_F$ given in (29) . We also have \be C_F \sin \pi
\nu = \left\{\begin{array} {cc} i \sqrt{f_1^2 + f_2^2 - f^2_1
f^2_2}& {\rm for} \,(e, e), \\
&\\
  \sqrt{ - f_1^2  + f_2^2(1 - f^2_1)} & {\rm for}
\,(e, m), \\
&\\
 \sqrt{f_1^2  + f_2^2 +  f^2_1 f^2_2} & {\rm for} \,(m, m).
\end{array}\right. \ee  Note that for an electric flux $0 < |f_i| < 1$
while for a magnetic flux $0 < |f_i| < \infty$ and so  $\nu = i
\nu_0$ with $0 < \nu_0 < \infty$ for case (e, e),  $\nu = \nu_0$
with $0 < \nu_0 < 1/2$ for case (m, m) but for (e, m), $\nu$ can be
either real or imaginary depending on whether the magnetic flux or
the electric flux dominates.  For (e, m), when \be |f_2| <
\frac{|f_1|}{\sqrt{1 - f_1^2}},\ee the corresponding $\nu$ is
imaginary, otherwise it will be real.  Also only for this case (as
well for (m, e) case), the corresponding amplitude can vanish with
non-vanishing fluxes, which signals the preservation of certain
number of corresponding spacetime supersymmetries. This actually
occurs at (now $\nu = 0$) \be f_2 =\pm \frac{f_1}{\sqrt{1 -
f_1^2}},\ee which gives rise to a quantized string coupling as \be
g_s = \frac{n_1}{n_2}\frac{|m_2|}{|m_1|}. \ee
 To validate our
computations, we need to have $g_s < 1$ which puts also constraint
on the respective two pairs of integers above. As will be shown in
the appendix, Eq. (40)  is precisely the condition for the
underlying system to  preserve also 1/4 of spacetime
supersymmetries.

 Our above amplitude can be greatly simplified if we make use of
the following identity as discussed in \cite{Lu:2009yx} for
$\theta$-functions \be 2 \, \theta^4_1 (\nu | \tau) = \theta_3 (2
\nu | \tau)\, \theta^3_3 (0 | \tau) - \theta_4 (2 \nu | \tau)\,
\theta^3_4 (0 | \tau) - \theta_2 (2 \nu | \tau)\, \theta^3_2 (0 |
\tau), \ee and it is given by  \bea \Gamma &=& \frac{2\, n_1 n_2 \,
V_{p + 1} \, C_F\, \sin \pi \nu}{ (8 \pi^2 \alpha')^{\frac{1 +
p}{2}}} \int_0^\infty \frac{d t}{t} \, e^{- \frac{Y^2}{2 \pi \alpha'
t}} \, t^{ - \frac{7 - p}{2}} \frac{1}{\eta^9 (it)} \frac{\theta^4_1
(\frac{\nu}{2} | it)}{\theta_1 (\nu |it)},\nn &=& \frac{2^4\, n_1
n_2 \, V_{p + 1} C_F \,\sin^4 {\frac{\pi \nu}{2}}}{ (8 \pi^2
\alpha')^{\frac{1 + p}{2}} } \int_0^\infty \frac{d t}{t} \, e^{-
\frac{Y^2}{2 \pi \alpha' t}} \, t^{ - \frac{7 - p}{2}} \nn
&\,&\qquad\qquad \qquad\qquad\quad\times \prod_{n = 1}^\infty
\frac{\left(1 - e^{i\pi \nu} |z|^{2n}\right)^4 \left(1 - e^{- i \pi
\nu} |z|^{2n}\right)^4}{\left(1 - |z|^{2n}\right)^6 \, \left(1 -
e^{2 i \pi \nu} |z|^{2n} \right)\, \left(1 - e^{- 2 i \pi \nu}
|z|^{2n}\right)},\eea where in the second equality we have made use
of explicit expressions for the Dedekind $\eta$-function and the
theta-function $\theta_1$ and in the above \be \sin^4 \frac{\pi
\nu}{2} = \frac{1}{4} \left(\cos \pi \nu - 1\right)^2 = \frac{1}{4}
(D_F - 1)^2.\ee

We now consider the large $Y$ limit of the amplitude (43) for $p \le
6$. This amounts to accounting for the massless-mode contribution of
closed string. Due to the exponential suppression of large $Y$, we
need only to keep the leading-order contributions of the following
in the integrand for large $t$,  \be \theta_1 (\nu | it) \rightarrow
2 e^{- \frac{\pi t}{4}} \sin \pi\nu, \qquad \theta_1 (\frac{\nu}{2}
| it) \rightarrow 2 e^{- \frac{\pi t}{4}} \sin \frac{\pi\nu}{2},
\qquad \eta (i t) \rightarrow e^{- \frac{\pi t}{12}}, \ee since now
$|z| = e^{- \pi t} \rightarrow 0$. So \bea \Gamma &\rightarrow&
\frac{2\, n_1 n_2 \, V_{p + 1} \, C_F\, \sin \pi \nu}{ (8 \pi^2
\alpha')^{\frac{1 + p}{2}}}\int_0^\infty \frac{d t}{t} \, e^{-
\frac{Y^2}{2 \pi \alpha' t}} \, t^{ - \frac{7 - p}{2}} \frac{1}{e^{-
\frac{3\pi t}{4}}} \frac{ 2^4\, e^{- \pi t}\, \sin^4
{\frac{\pi\nu}{2}}}{2\, e^{- \frac{\pi t}{4}} \sin\pi \nu},\nn &=\,&
\frac{2^4 \, n_1 n_2 \, V_{p + 1}\, C_F \,\sin^4\frac{\pi\nu}{2} }{
(8 \pi^2 \alpha')^{\frac{1 + p}{2}}} \int_0^\infty \frac{d t}{t} \,
e^{- \frac{Y^2}{2 \pi \alpha' t}} \, t^{ - \frac{7 - p}{2}}, \nn
&=\,& \frac{2^4 \, n_1 n_2 \, V_{p + 1}\, C_F
\,\sin^4\frac{\pi\nu}{2} }{ (8 \pi^2 \alpha')^{\frac{1 + p}{2}}}
\left(\frac{2\pi\alpha'}{Y^2}\right)^{\frac{7 - p}{2}} \Gamma
\left(\frac{7 - p}{2}\right), \nn &=\,& \frac{C(m_1, n_1; m_2,
n_2)}{Y^{7 - p}}\eea where \be C (m_1, n_1; m_2, n_2) = \frac{c_p^2
\, V_{p + 1} U (m_1, n_1; m_2, n_2)}{(7 - p) \, \Omega_{8 - p}}. \ee
In the above, $(7 - p)\Omega_{8 - p} = 4 \pi \pi^{(7 - p)/2}/\Gamma
((7 - p)/2)$ with $\Omega_q$ the volume of unit q-sphere and \bea U
(m_1, n_1; m_2, n_2) &\equiv&  4 \, n_1 n_2 C_F \sin^4 \frac{\pi
\nu}{2}, \nn & =& \left\{\begin{array} {cc}
\frac{\left(n_1 n_2 - g_s^2 \Omega_{ee}\right)^2}{g_s^2 \,\Omega_{ee}} & {\rm for} \,(e, e), \\
&\\
  \frac{\left(g_s n_2 \, \triangle^{1/2}_{e (m_1, n_1)} - n_1 \triangle^{1/2}_{m (m_2, n_2)}\right)^2}{g_s\, \Omega_{e m}}  & {\rm for}
\,(e, m), \\
&\\
 \frac{\left(n_1 n_2 - \Omega_{mm}\right)^2}{\Omega_{mm}} & {\rm for} \,(m,
 m),
\end{array}\right. \eea
where $\Omega_{ee} = \triangle^{1/2}_{e(m_1, n_1)}
\triangle^{1/2}_{e (m_2, n_2)},\, \Omega_{em} =
\triangle^{1/2}_{e(m_1, n_1)} \triangle^{1/2}_{m (m_2, n_2)}$ and
$\Omega_{mm} = \triangle^{1/2}_{m(m_1, n_1)} \triangle^{1/2}_{m
(m_2, n_2)}$ with $\triangle_{e(m_i, n_i)}$ and $\triangle_{m (m_i,
n_i)}$ defined in (23) and (25), respectively. For a non-vanishing
flux, either electric or magnetic, i.e, $m_i \neq 0$ ($i = 1, 2$),
the above $U (m_1, n_1; m_2, n_2)$ can vanish only for the case (e,
m) (or (m, e)) and if this occurs, the corresponding amplitude as
well as its large separation limit vanishes. The condition for this
to occur is exactly the same as the one given in (41).

  We will have $U(m_1, n_1; m_2, n_2) >
 0$ in all cases considered above if the string coupling constant is not
 quantized as given in (41) for the case of (e, m).
  Note that each numerator in the infinite product in the
integrand in the second equality of (43) can be re-expressed as \be
\left(1 - e^{i\pi \nu} |z|^{2n}\right)^4 \left(1 - e^{- i\pi \nu}
|z|^{2n}\right)^4 = \left(1 - 2 \cos\pi\nu\, |z|^{2n} +
|z|^{4n}\right)^4 > 0, \ee so the sign of the interaction amplitude
will depend on that of the factor in  each denominator in the
infinite product in the integrand \be \left(1 - e^{2 i \pi \nu}
|z|^{2n}\right) \left(1 - e^{- 2 i\pi \nu} |z|^{2n}\right)= \left(1
- 2 \cos{2\pi\nu}\, |z|^{2n} + |z|^{4n}\right), \ee which is always
positive for the case of (m, m) and the case of (e, m) when (39) is
not satisfied, respectively. In other words, for the later case, the
magnetic flux plays at least the equally important role as the
electric flux and now $\nu$ is real. Then the corresponding
interaction amplitude in each of the above two cases is greater than
zero and is solely determined by the positiveness of $U (m_1, n_1;
m_2, n_2)$. In this aspect it shares the same feature as its long
distance interaction, reflecting the attractive nature of the
interaction. While this factor is still positive for large $t$ but
it can be negative for small $t$ for the case of (e, e) and the case
of (e, m) when (39) is satisfied, respectively. In other words, for
the (e, m) case, the electric flux now plays a dominant role and
$\nu$ is now imaginary. For either of the present two cases, while
the long distance interaction amplitude is again greater than zero
(implying also an attractive interaction ) and is also solely
determined by the positiveness of the corresponding $U (m_1, n_1;
m_2, n_2)$, the sign of the small separation amplitude
(corresponding to small $t$ contribution) is uncertain in the
present representation of integration variable $t$ since even with
the factor in (50) less than zero, the sign of the product of
infinite number of such factors in the integrand remains indefinite.
So one  expects some interesting physics to appear in this case for
small $t$.

The small $t$ contribution to the amplitude mainly concerns about
the physics for small separation $Y$. The appropriate frame for
describing the underlying physics as well as the analytic structure
as a function of the separation in the short cylinder limit $t
\rightarrow 0$ is in terms of an annulus, which can be achieved by
the Jacobi transformation $t \rightarrow t' = 1/t$. This is also
stressed in \cite{Douglas:1996yp} that the lightest open string
modes now contribute most and the open string description is most
relevant. So in terms of the annulus variable $t'$, noting \bea \eta
(\tau) &=& \frac{1}{\left(- i \tau\right)^{1/2}} \eta \left(-
\frac{1}{\tau}\right),\nn \theta_1 (\nu|\tau) &=& i \frac{\,e^{- i
\pi \nu^2/\tau}}{\left(- i \tau\right)^{1/2}} \theta_1
\left(\left.\frac{\nu}{\tau}\right|- \frac{1}{\tau}\right),\eea the
amplitude  in (43) can now be reexpressed as \bea \Gamma &=& - i
\frac{U(m_1, n_1; m_2, n_2)\, V_{p + 1} }{ 2 (8 \pi^2
\alpha')^{\frac{1 + p}{2}}} \frac{\sin \pi \nu}{\sin^4 \frac{\pi
\nu}{2}}\,  \int_0^\infty \frac{d t'}{t'} \, e^{- \frac{Y^2 t'}{2
\pi \alpha' }} \, t'^{ \frac{1 - p}{2}} \frac{1}{\eta^9 (it')}
\frac{\theta^4_1 (\frac{- i\nu t'}{2} | it')}{\theta_1 (-i \nu
t'|it')},\nn &=& - i \frac{4 \,U(m_1, n_1; m_2, n_2)\, V_{p + 1} }{
(8 \pi^2 \alpha')^{\frac{1 + p}{2}}} \frac{\sin \pi \nu}{\sin^4
\frac{\pi \nu}{2}}\, \int_0^\infty \frac{d t'}{t'} \, e^{- \frac{Y^2
t'}{2 \pi \alpha' }} \, t'^{ \frac{1 - p}{2}} \frac{\sin^4
\left(\frac{- i \pi \nu t'}{2}\right)}{\sin \left(- i \pi \nu
t'\right)} \nn &\,& \qquad\qquad \quad\times \prod_{n = 1}^\infty
\frac{\left(1 - e^{\pi \nu t'} |z|^{2n}\right)^4 \left(1 - e^{-\pi
\nu t'} |z|^{2n}\right)^4}{\left(1 - |z|^{2n}\right)^6 \, \left(1 -
e^{2 \pi \nu t'} |z|^{2n} \right)\, \left(1 - e^{- 2  \pi \nu t'}
|z|^{2n}\right)},\eea where we have made use of the expression for
$U (m_1, n_1; m_2, n_2)$ given in (48) and now $|z| = e^{- \pi t'}$.
We follow \cite{Green:1996um, Lu:2009yx} to discuss the underlying
analytic structure and the possible associated physics of the
amplitude of (52). For the case when $\nu$ is real as
 mentioned above,  we limit ourselves to the interesting non-BPS
amplitude, i.e., $\nu = \nu_0$ with $0 < \nu_0 < 1/2$, and for this
the above amplitude is purely real and has no singularities unless
$Y \leq \pi \sqrt{2 \nu \alpha'}$, i.e. on the order of string
scale, for which the integrand is dominated by, in the short
cylinder limit $t' \rightarrow \infty$, \be \lim_{t' \rightarrow
\infty} \frac{e^{- \frac{Y^2 t'}{2\pi \alpha'}} \theta_1 (- i \pi
\nu t'/2 |i t')}{i\, \eta (it') \theta_1 (- i \pi \nu t'| it')} \sim
\lim_{t' \rightarrow \infty} \frac{e^{- \frac{Y^2 t'}{2\pi \alpha'}}
\sin^4(- i \pi \nu t'/2)}{i\, \sin (- i \pi\nu t')} \sim
\lim_{t'\rightarrow \infty} e^{- \frac{t'}{2\pi\alpha'}(Y^2 - 2\pi^2
\nu \alpha')}. \ee The contribution of the annulus to the vacuum
amplitude (energy) should be real if the integrand in (52) have no
simple poles on the positive $t'$-axis  since the imaginary part of
the amplitude is given by the sum of residues at the poles times
$\pi$ due to the integration contour passing to the right of all
poles as dictated by the proper definition of the Feynman
propagator\cite{Bachas:1992bh}. In the present case, the amplitude
appears purely real and there are no simple poles on the positive
$t'$-axis, therefore giving zero imaginary amplitude, i.e., zero
pair-production (absorptive) rate of open strings, which is
consistent with the conclusion reached in \cite{Schwinger:1951nm} in
quantum field theory context and also pointed out in a similar
context in \cite{Bachas:1992zr, Lu:2009yx}. When $Y \leq \pi \sqrt{2
\nu_0 \alpha'}$, i.e., on the order of string scale, the amplitude
diverges as indicated in (56) and this happens in a similar fashion
as in the case of a brane/antibrane system as discussed in
\cite{Banks:1995ch, Lu:2007kv} but now caused by the presence of
dominant magnetic flux or fluxes. The appearance of the divergent
amplitude indicates the breakdown of the calculations, signalling
the onset of tachyonic instability caused by the dominant magnetic
flux or fluxes\footnote{With $0 < \nu = \nu_0 < 1/2$, in addition to
the evidence given in the text, that the open string tachyon mode
appears to arise is also indicated from the leading term $e^{\pi \nu
t'}$, which diverges in the short cylinder limit $t' \rightarrow
\infty$, in the expansion of the $\theta$-functions and
$\eta$-function in (52) in the open string channel. Note also that
in the case of (e, m), $\nu$ can be zero when the string coupling
satisfies (41) and this divergent term, therefore the tachyon mode,
then disappears and this is entirely consistent with the fact that
the amplitude also vanishes, signalling the underlying system being
BPS and preserving certain number of spacetime supersymmetries. }
and the relaxation of the system to form a new non-threshold bound
state. However,  the detail of this requires further dynamical
understanding.

Let us move to the case when the electric flux or fluxes are
dominant in a sense mentioned earlier. We have now $\nu = i \nu_0$
with $0 < \nu_0 < \infty$ ($\nu_0 = 0$ corresponds to BPS case and
is not considered here). The amplitude (52) is now \bea \Gamma  &=&
 \frac{4 \,U(m_1, n_1; m_2, n_2)\, V_{p + 1} }{ (8 \pi^2
\alpha')^{\frac{1 + p}{2}}} \frac{\sinh \pi \nu_0}{\sinh^4 \frac{\pi
\nu_0}{2}}\, \int_0^\infty \frac{d t'}{t'} \, e^{- \frac{Y^2 t'}{2
\pi \alpha' }} \, t'^{ \frac{1 - p}{2}} \frac{\sin^4 \left(\frac{
\pi \nu_0 t'}{2}\right)}{\sin \left( \pi \nu_0 t'\right)} \nn &\,&
\qquad\qquad \quad\times \prod_{n = 1}^\infty \frac{\left(1 - e^{i
\pi \nu_0 t'} |z|^{2n}\right)^4 \left(1 - e^{- i\pi \nu_0 t'}
|z|^{2n}\right)^4}{\left(1 - |z|^{2n}\right)^6 \, \left(1 - e^{ 2 i
\pi \nu_0 t'} |z|^{2n} \right)\, \left(1 - e^{- 2 i \pi \nu_0
t'}|z|^{2n}\right)}.\eea Exactly the same as the cases discussed in
\cite{Lu:2009yx, Green:1996um}, the above integrand has also an
infinite number of simple poles on the positive real $t'$-axis at
$t' = (2 k + 1)/\nu_0$ with $k = 0, 1, 2, \cdots$. This leads to an
imaginary part of the amplitude, which is given as the sum over the
residues of the poles as described in \cite{Bachas:1992bh,
Bachas:1995kx}. Therefore the rate of pair production of open
strings per unit worldvolume  in the present context is \bea {\cal
W} &\equiv& - \frac{2 {\rm Im} \Gamma}{V_{p + 1}},\nn &=& \frac{8
\,U(m_1, n_1; m_2, n_2) }{\nu_0 (8 \pi^2 \alpha')^{\frac{1 + p}{2}}}
\frac{\sinh \pi \nu_0}{\sinh^4 \frac{\pi \nu_0}{2}}\,\sum_{k =
0}^\infty \left(\frac{\nu_0}{2 k + 1}\right)^{\frac{1 + p}{2}} e^{-
\frac{(2 k + 1) Y^2 }{2 \pi \nu_0 \alpha' }} \, \prod_{n = 1}^\infty
\left(\frac{1 + e^{- 2n\pi (2 k + 1)/\nu_0}}{ 1 - e^{- 2 n (2 k +
1)\pi/\nu_0 }}\right)^8, \nn &=& \frac{32 n_1 n_2 \tanh \pi \nu_0
}{\nu_0 (8 \pi^2 \alpha')^{\frac{1 + p}{2}}}\,\sum_{k = 0}^\infty
\left(\frac{\nu_0}{2 k + 1}\right)^{\frac{1 + p}{2}} e^{- \frac{(2 k
+ 1) Y^2 }{2 \pi \nu_0 \alpha' }} \, \prod_{n = 1}^\infty
\left(\frac{1 + e^{- 2n (2 k + 1)\pi/\nu_0}}{ 1 - e^{- 2 n (2 k +
1)\pi/\nu_0 }}\right)^8,\nn\eea where we have used $U (m_1, n_1;
m_2, n_2) = 4\, n_1 n_2 C_F \sinh^4 \pi\nu_0/2$ in the present
context and  $\nu_0$ can be determined from \be \tanh \pi \nu_0 =
\left\{\begin{array} {cc} \frac{\sqrt{g_s^2 m_1^2 m_2^2 + n_1^2
m_2^2 + n_2^2 m_1^2}}{g_s \,
\Omega_{ee}} & {\rm for} \,(e, e), \\
&\\
 \frac{\sqrt{ m_1^2 n_2^2 g_s^2 - n_1^2 m_2^2}}{g_s n_2 \triangle^{1/2}_{e (m_1, n_1)}} & {\rm for}
\,(e, m), \\
\end{array}\right. \ee
where $\triangle_{e(m_i, n_i)}$ is defined in (23) with $i = 1, 2$
and $\Omega_{ee} = \triangle^{1/2}_{e(m_1,n_1)}
\triangle^{1/2}_{e(m_2, n_2)}$ as defined earlier. Also in the
above, the condition (39) for the case of (e, m) needs to be
satisfied. In the present context, it is $g_s
> n_1 |m_2|/(n_2 |m_1|)$.
  Note that the above rate is
suppressed by the brane separation and the integer $k$ but increases
with the value of $\nu_0$ which is expected. Let us consider $\nu_0
\rightarrow 0$ and $\nu_0 \rightarrow \infty$ limits for each case
considered above.  The former limit corresponds to the near extremal
limit which requires both electric fluxes to be small or
equivalently $n_i \gg g_s m_i$ with $i = 1, 2$ for the case of (e,
e) and  $(g_s n_2 |m_1| - n_1 |m_2|) \rightarrow 0^+$ for the case
of (e, m). In either of these two cases, $\nu_0 \rightarrow 0$ and
$\tanh\pi\nu_0 \rightarrow \pi \nu_0$. The rate is now approximated
well by the leading $k = 0$ term as \be {\cal W} \approx 32 n_1
n_2\,\pi\left(\frac{\nu_0}{8\pi^2 \alpha'}\right)^{\frac{1 + p}{2}}
 e^{- \frac{Y^2}{2\pi \nu_0 \alpha' }}, \ee vanishing small
as expected. The $\nu_0 \rightarrow \infty$ limit requires that the
electric flux or fluxes all reach their respective critical field
limit in  either case considered here. In addition, we need $g_s\,
|m_1| /n_1 \gg |m_2|/n_2$ for the case of (e, m). Then each term in
the summation of (55) diverges and so does the rate, signalling also
an instability as mentioned in a similar context in
\cite{Porrati:1993qd}.

\subsection{Two fluxes without common index}
We now discuss the cases when the two non-vanishing worldvolume
constant fluxes $(\hat F_1)_{\alpha\beta}$ and $(\hat
F_2)_{\gamma\delta}$
 specified at the beginning of this section share  no common
index, i.e., $\alpha, \beta \neq \gamma, \delta$. So we have only
three cases to consider: 1) (e, m), 2) (m, e) and 3) (m, m). For the
former two cases, we need $p \geq 3$ for the spatial dimensions of
Dp branes in the non-threshold bound states while for the later
case, we need $p \geq 4$. If $(\hat F_1)_{\alpha\beta}$ is an
electric flux as specified in (18), then we have the case 1) above
with $(\hat F_2)_{\gamma\delta}$ a magnetic flux.  We choose then
its only two non-vanishing components $(\hat F_2)_{cd} = - (\hat
F_2)_{ dc} = - f_2$ at two given spatial indices $c, d$ with the
constraint $c < d$ and $c \neq 1$. If $(\hat F_1)_{\alpha\beta}$ is
a magnetic flux as specified in (20), we can have either case 2)
above with the only two non-vanishing electric flux components
$(\hat F_2)_{ 0a} = - (\hat F_2)_{ a0} = - f_2$ with $a \neq p - 1,
p$ or the case 3) with the only two non-vanishing magnetic
components $(\hat F_2)_{ cd} = - (\hat F_2)_{ dc} = - f_2$ with now
the spatial indices $c < d$ and $c, d \neq p - 1, p$. We have then
the various matrix elements specified in (17) as
 \bea A^X &=& C_F \, V_{p + 1}\, e^{- \frac{Y^2}{2\pi
\alpha' t}} \left(2\pi^2 \alpha'\, t\right)^{- \frac{9 - p}{2}}
\prod_{n = 1}^\infty \frac{1}{(1 - |z|^{2n})^6}\prod_{j = 1}^2
\frac{1}{(1 - \lambda_j |z|^{2n})(1 - \lambda_j^{-1} |z|^{2n})},\nn
A^{bc} &=& |z|^{-2} \prod_{n = 1}^\infty (1 - |z|^{2n})^2,\eea for
both NS-NS and R-R sectors, \bea &&A_{\rm NS}^{\beta\gamma} (\eta',
\eta) = |z| \prod_{n = 1}^{\infty} \frac{1}{(1 + \eta' \eta\,
|z|^{2n - 1})^2},\nn &&A^\psi_{\rm NS} = \prod_{n = 1}^\infty (1 +
\eta' \eta \,|z|^{2n - 1})^6 \prod_{j = 1}^2 (1 + \eta' \eta
\,\lambda_j |z|^{2n - 1}) (1 + \eta' \eta\,\lambda_j^{-1}\, |z|^{2n
- 1}), \eea for NS-NS sector, and \be A^{\beta\gamma}_{\rm R}
(\eta', \eta) A^\psi_{\rm R} (\eta', \eta) = - 2^4 \,|z|^2\, D_F\,
\delta_{\eta'\eta, + }\prod_{n = 1}^\infty ( 1 + |z|^{2n})^4
\prod_{j = 1}^2 (1 + \lambda_j\, |z|^{2n}) (1 + \lambda_j^{- 1} \,
|z|^{2n} ), \ee for the R-R sector. Note that we have $|z| = e^{-
\pi t}$ above and again in (60) we follow the prescription given in
\cite{Billo:1998vr,Di Vecchia:1999rh} not to separate the
contributions from matter fields $\psi^\mu$ and superghosts in the
R-R sector in order to avoid the complication due to the respective
zero modes.  Also in the above\footnote{By the same token, the (m,
e) case can be similarly discussed, therefore not repeated in what
follows for simplicity.}, we have \be D^{-1}_F  = C_F =
\left\{\begin{array} {cc}  \sqrt{(1 - f_1^2)(1 + f_2^2)} & {\rm for}
\,(e, m), \\
&\\
  \sqrt{(1 + f_1^2)(1 + f_2^2)} & {\rm for} \,(m, m),\\
\end{array}\right. \ee

and \be \lambda_1+ \lambda_1^{- 1} =  2 (2 D_{F_1}^2 - 1) =
\left\{\begin{array} {cc}  2 \frac{1 + f_1^2}{1 - f_1^2} & {\rm for}
\,(e, m), \\
&\\
2 \frac{1 - f_1^2}{1 + f_1^2} & {\rm for} \,(m, m),
\end{array}\right. \ee
\be \lambda_2+ \lambda_2^{- 1} =  2 (2 D_{F_2}^2 - 1) =
\left\{\begin{array} {cc}  2 \frac{1 - f_2^2}{1 + f_2^2} & {\rm for}
\,(e, m), \\
&\\
 2 \frac{1 - f_2^2}{1 + f_2^2} & {\rm for} \,(m, m),
\end{array}\right. \ee where $D_F = D_{F_1} D_{F_2}$.

With the above matrix elements and by using (15) and (16), we can
have the amplitude in the NS-NS sector using (13) with the GSO
projected boundary state in the NS-NS sector defined in (1) as \bea
\Gamma_{\rm NS} &=& \frac{n_1 n_2 \, V_{p + 1}\, C_F}{2 (8 \pi^2
\alpha')^{\frac{1 + p}{2}}} \int_0^\infty \frac{d t}{t} \, e^{-
\frac{Y^2}{2 \pi \alpha' t}} \, t^{ - \frac{7 - p}{2}} \nn &\,&
\times |z|^{ - 1} \left[\prod_{n = 1}^\infty \frac{(1 + |z|^{2n -
1})^4}{(1 - |z|^{2n})^4} \prod_{j = 1}^2\frac{(1 + \lambda_j\,
|z|^{2n - 1})(1 + \lambda_j^{- 1} \,|z|^{2n - 1})} {(1 - \lambda_j\,
|z|^{2n})(1 - \lambda_j^{- 1} \, |z|^{2n})}\right.\nn
&&\qquad\qquad\left. - \prod_{n = 1}^\infty \frac{(1 - |z|^{2n -
1})^4} {(1 - |z|^{2n})^4}\prod_{j = 1}^2 \frac{(1 - \lambda_j\,
|z|^{2n - 1})(1 -\lambda_j^{- 1} \,|z|^{2n - 1})}{(1 - \lambda_j\,
|z|^{2n})(1 - \lambda_j^{- 1} \, |z|^{2n})} \right],\eea and
similarly we have the amplitude in the R-R sector using the GSO
projected boundary state given in (2) as \bea \Gamma_{R} &=&  -
\frac{ 8\, n_1 n_2\, V_{p + 1}}{ (8 \pi^2 \alpha')^{\frac{1 +
p}{2}}} \int_0^\infty \frac{d t}{t}\,e^{- \frac{Y^2}{2 \pi \alpha'
t}} \, t^{ - \frac{7 - p}{2}} \nn && \times \prod_{n =
1}^\infty\frac{(1 + |z|^{2n})^4} {(1 - |z|^{2n})^4}\prod_{j = 1}^2
\frac{(1 + \lambda_j\, |z|^{2n})(1 + \lambda_j^{- 1} \,|z|^{2n})}{(1
- \lambda_j\, |z|^{2n})(1 - \lambda_j^{- 1} \, |z|^{2n})} .\eea In
obtaining the above amplitudes, we have made use of equations (33)
and (35) and some cautions mentioned in the previous subsection
below (35) also apply and will not be repeated here.

The total amplitude is then \bea  \Gamma &=& \Gamma_{\rm NS} +
\Gamma_{\rm R} \nn &=& \frac{n_1 n_2 \, V_{p + 1}\, C_F}{2 (8 \pi^2
\alpha')^{\frac{1 + p}{2}}} \int_0^\infty \frac{d t}{t} \, e^{-
\frac{Y^2}{2 \pi \alpha' t}} \, t^{ - \frac{7 - p}{2}}\nn &\,&\times
\left\{|z|^{ - 1} \left[\prod_{n = 1}^\infty \frac{(1 + |z|^{2n -
1})^4}{(1 - |z|^{2n})^4} \prod_{j = 1}^2\frac{(1 + \lambda_j\,
|z|^{2n - 1})(1 + \lambda_j^{- 1} \,|z|^{2n - 1})} {(1 - \lambda_j\,
|z|^{2n})(1 - \lambda_j^{- 1} \, |z|^{2n})}\right.\right.\nn
&\,&\qquad\qquad \left. - \prod_{n = 1}^\infty \frac{(1 - |z|^{2n -
1})^4} {(1 - |z|^{2n})^4}\prod_{j = 1}^2 \frac{(1 - \lambda_j\,
|z|^{2n - 1})(1 -\lambda_j^{- 1} \,|z|^{2n - 1})}{(1 - \lambda_j\,
|z|^{2n})(1 - \lambda_j^{- 1} \, |z|^{2n})} \right]\nn && \qquad
\left. - 2^4 D_F \prod_{n = 1}^\infty\frac{(1 + |z|^{2n})^4} {(1 -
|z|^{2n})^4}\prod_{j = 1}^2 \frac{(1 + \lambda_j\, |z|^{2n})(1 +
\lambda_j^{- 1} \,|z|^{2n})}{(1 - \lambda_j\, |z|^{2n})(1 -
\lambda_j^{- 1} \, |z|^{2n})} \right\} ,\eea where the structure
looks a bit different from the one given in the previous subsection
and in \cite{Lu:2009yx} for which the two fluxes share at least one
common direction. As we will show, the previous structure for
amplitudes is just a special case of the present one. The above
amplitude can be re-expressed in a nice form in terms of various
$\theta$-functions and the Dedekind $\eta$-function with their
standard definitions as mentioned in the previous subsection. We
then have \bea \Gamma &=& \frac{2 \,n_1 n_2 \, V_{p + 1} \,  \tan
\pi \nu_1 \,\tan \pi\nu_2}{ (8 \pi^2 \alpha')^{\frac{1 + p}{2}}}
\int_0^\infty \frac{d t}{t} \, e^{- \frac{Y^2}{2 \pi \alpha' t}} \,
t^{ - \frac{7 - p}{2}} \nn &\,&\times \frac{1}{\eta^6 (it)}
\left[\frac{\theta_3 (\nu_1|it)\, \theta_3 (\nu_2|it) \,\theta^2_3
(0|it)}{\theta_1 (\nu_1 |it)\,\theta_1 (\nu_2 | it)} -
\frac{\theta_4 (\nu_1 | it)\,\theta_4 (\nu_2 | it)\, \theta^2_4 (0 |
it)} {\theta_1 (\nu_1 | it)\theta_1 (\nu_2| it)}\right.\nn &&\left.
\qquad\qquad - \frac{\theta_2 (\nu_1 | it) \,\theta_2
(\nu_2|it)\,\theta^2_2 ( 0 | it)}{\theta_1 (\nu_1 | it)\, \theta_1
(\nu_2 | it)} \right],\eea where we have defined $\lambda_j =
e^{2\pi i \nu_j}$ and used the fact $\cos\pi \nu_j = D_{F_j}$ which
can be obtained from $\lambda_j + \lambda_j^{-1} = 2 (2 D_{F_j}^2 -
1)$ as given in (62) and (63)  with $D_{F_j} = 1/\sqrt{1 - f_j^2}$
for an electric flux and $D_{F_j} = 1/\sqrt{1 + f_j^2}$ for a
magnetic flux for $j = 1, 2$, respectively.  We also have \be \tan
\pi \nu_j = \left\{\begin{array} {cc} i |f_j| & {\rm for} \,{\rm an}\, {\rm electric}\,{\rm flux}, \\
&\\
 |f_j|  & {\rm for}\, {\rm a} \, {\rm magnetic}\, {\rm flux}, \\
\end{array}\right. \ee where the subscript index $j = 1, 2$. Note that for an electric flux $0 < |f_j| < 1$
while for a magnetic flux $0 < |f_j| < \infty$ and so  $\nu_j = i
\nu_{j0}$ with $0 < \nu_{j0} < \infty$ for an electric flux and
$\nu_j = \nu_{j0}$ with $0 < \nu_{j0} < 1/2$ for a magnetic flux.
 The above amplitude can be
greatly simplified if the following identity for $\theta$-functions
is employed\footnote{This identity can be obtained from the general
one (iv) on page 468 given in \cite{whittaker-watson}. The notations
there for various $\theta$-functions are $\theta_r (z) \equiv
\theta_r (z |\tau)$ with $r = 1, 2, 3, 4$. In obtaining (69) in the
text, we need to make choices for the variables as $y' = 0, z' = 0,
w' = x + y, x' = - x + y$ and set $x = (\nu_1 - \nu_2)/2, y = (\nu_1
+ \nu_2)/2$.} \bea 2 \,\theta^2_1 \left(\left.\frac{\nu_1 -
\nu_2}{2}\right|\tau\right)\, \theta^2_1 \left(\left.\frac{\nu_1 +
\nu_2}{2}\right|\tau\right) &=& \theta_3 (\nu_1|\tau) \theta_3
(\nu_2 |\tau) \theta^2 (0 |\tau) - \theta_4 (\nu_1 |\tau) \theta_4
(\nu_2 |\tau) \theta^2_4 (0 |\tau) \nn &\,&- \theta_2 (\nu_1
|\tau)\theta_2 (\nu_2 | \tau) \theta_2 (0 | \tau),\eea and the
amplitude becomes \bea \Gamma &=& \frac{4 \,n_1 n_2 \, V_{p + 1} \,
\tan \pi \nu_1 \,\tan \pi\nu_2}{ (8 \pi^2 \alpha')^{\frac{1 +
p}{2}}} \int_0^\infty \frac{d t}{t} \, e^{- \frac{Y^2}{2 \pi \alpha'
t}} \, t^{ - \frac{7 - p}{2}} \frac{1}{\eta^6 (it)} \frac{\theta^2_1
\left(\left.\frac{\nu_1 - \nu_2}{2}\right|it\right)\, \theta^2_1
\left(\left.\frac{\nu_1 + \nu_2}{2}\right|it\right)}{\theta_1
(\nu_1|it) \theta_1 (\nu_2 |it)},\nn
 &=& \frac{2^4 \,n_1 n_2 \, V_{p + 1} \,C_F
\sin^2 \frac{\pi(\nu_1 - \nu_2)}{2} \,\sin^2\frac{\pi(\nu_1 +
\nu_2)}{2}}{ (8 \pi^2 \alpha')^{\frac{1 + p}{2}}} \int_0^\infty
\frac{d t}{t} \, e^{- \frac{Y^2}{2 \pi \alpha' t}} \, t^{ - \frac{7
- p}{2}}\nn &\,&\qquad \times \prod_{n = 1}^\infty \frac{1}{(1 -
|z|^{2n})^4} \prod_{j = 1}^2 \frac{(1 - e^{\pi i(\nu_1 + (-)^j
\nu_2)} |z|^{2n})^2 (1 - e^{- \pi i (\nu_1 + (-)^j \nu_2)}
|z|^{2n})^2}{(1 - e^{2\pi i \nu_j} |z|^{2n})(1 - e^{- 2\pi i \nu_j}
|z|^{2n})},\eea where in the second equality we have used the
explicit expression for $\theta_1 (\nu|\tau)$ and $C_F^{-1} = D_F =
D_{F_1} D_{F_2}$ with $D_{F_j} = \cos\pi\nu_j$. One can check easily
with the known properties of $\theta_1 (\nu|\tau)$ that both (70)
and (67) will reduce their basic structures to their corresponding
ones obtained in the previous subsection or in \cite{Lu:2009yx}
where the two worldvolume fluxes share at least one common direction
if we set either
 $\nu_1 \,{\rm or}\, \nu_2 \rightarrow 0$ in (70) and (67).
 Therefore the basic structure of either (70) or (67) is more
 general than their respective previous correspondence just mentioned. For later purpose,
 let us define the following quantity similar to (48) as \bea U
(m_1, n_1; m_2, n_2) &\equiv&  4 \, n_1 n_2 C_F \sin^2 \frac{\pi
(\nu_1 - \nu_2)}{2} \sin^2\frac{\pi (\nu_1 + \nu_2)}{2} = n_1 n_2
\frac{(D_{F_1} - D_{F_2})^2}{D_{F_1} D_{F_2}}, \nn & =&
\left\{\begin{array} {cc}
\frac{\left(n_1 n_2 - g_s \Omega_{em}\right)^2}{g_s \,\Omega_{em}} & {\rm for} \,(e, m), \\
&\\
 \frac{\left(n_1 \triangle^{1/2}_{m(m_2, n_2)}  - n_2 \triangle^{1/2}_{m (m_1, n_1)}\right)^2}{\Omega_{mm}} & {\rm for} \,(m,
 m),
\end{array}\right. \eea where $\Omega_{em} =
\triangle^{1/2}_{e(m_1, n_1)} \triangle^{1/2}_{m(m_2, n_2)}$ and
$\Omega_{mm} = \triangle^{1/2}_{m(m_1, n_1)}
 \triangle^{1/2}_{m(m_2, n_2)}$  as before with $\triangle_{e(m_j, n_j)}$ and
$\triangle_{m (m_j, n_j)}$ defined in (23) and (25), respectively.
One can check easily that only for the (m, m) case above, $U (m_1,
n_1; m_2, n_2)$ can vanish and this occurs at $|m_1|/n_1 =
|m_2|/n_2$ ($n_1 n_2 > 0$) or $f_1 = \pm f_2$, giving a vanishing
amplitude, an indication of preservation of certain number of
spacetime supersymmetries. This is interesting and a bit
counterintuitive, and it indicates that when the magnetic flux in
one non-threshold bound state shares no common direction with the
magnetic flux in the other non-threshold bound state and when their
magnitude is the same, then the force acting between the two
cancels. As will be shown in the appendix, the above condition is
precisely the one for this system to preserve also 1/4 of spacetime
supersymmetries.

Following what we did in the previous subsection, the large
separation amplitude can be obtained from (70) and is \be \Gamma =
\frac{C(m_1, n_1; m_2, n_2)}{Y^{7 - p}},\ee where \be C (m_1, n_1;
m_2, n_2) = \frac{c_p^2 \, V_{p + 1}\, U (m_1, n_1; m_2, n_2)}{(7 -
p) \Omega_{8 - p}},\ee with $U (m_1, n_1; m_2, n_2)$ now given in
(71).

Note that if the condition $|m_1|/n_1 = |m_2|/n_2$ ($n_1 n_2 > 0$),
giving rise to $U (m_1, n_1; m_2, n_2) = 0$, for the case of (m, m)
mentioned above is excluded, we have then $U (m_1, n_1; m_2, n_2) >
0$  for all cases considered in this subsection. Note also the
following factor in the numerator in the infinite product in the
integrand in the second equality in (70) \bea &&\prod_{j = 1}^2 (1 -
e^{\pi i(\nu_1 + (-)^j \nu_2)} |z|^{2n})^2 (1 - e^{- \pi i (\nu_1 +
(-)^j \nu_2)} |z|^{2n})^2 = \nn && \left[(1 + |z|^{4n})^2 + 2
|z|^{4n} (\cos 2\pi \nu_1 + \cos2\pi \nu_2) - 4 |z|^{2n} (1 +
|z|^{4n}) \cos\pi\nu_1 \cos\pi\nu_2 \right]^2 > 0,\qquad \eea so the
sign of the amplitude is again determined by the following factor in
the denominator in the infinite product in the integrand \bea
&&\prod_{j = 1}^2 (1 - e^{2\pi i \nu_j} |z|^{2n})(1 - e^{- 2\pi i
\nu_j} |z|^{2n}) = \nn && (1 - 2  |z|^{2n} \cos2\pi\nu_1 +
|z|^{4n})(1 - 2 |z|^{2n} \cos2\pi \nu_2 + |z|^{4n}),\eea which is
positive for the case of (m, m) for which both $\nu_1$ and $\nu_2$
are real and  for large $t$ for the remaining case but can be
negative for small t for this case since now  $\nu_1$  is imaginary.
Therefore the interaction amplitude is positive for the case of (m,
m) once again as expected, reflecting the attractive nature of the
interaction between BS1 and BS2 in the present case. For the case of
(e, m) while the large separation amplitude is still positive and
the corresponding interaction is attractive, the small separation
amplitude is once again uncertain for the same reason mentioned in
the previous subsection in a similar situation. We again expect
interesting physics to arise in the small $t$ limit for these two
cases to which we will turn next.

The best picture to study the small $t$ physics is in terms of open
string description \cite{Douglas:1996yp} and this can be realized by
the transformation of integration variable $t \rightarrow t' = 1/t$
which converts the closed string cylinder diagram to the open string
annulus diagram. So in terms of the annulus variable $t'$, with
(51), the amplitude (70) is now \bea \Gamma &=& - \frac{4 \,n_1 n_2
\, V_{p + 1} \, \tan \pi \nu_1 \,\tan \pi\nu_2}{ (8 \pi^2
\alpha')^{\frac{1 + p}{2}}} \int_0^\infty \frac{d t'}{t'} \, e^{-
\frac{Y^2 t'}{2 \pi \alpha' }} \, t'^{  \frac{3 - p}{2}}
\frac{\theta^2_1(- i\frac{ \nu_1 - \nu_2}{2} t'|it')\, \theta^2_1 (
- i \frac{\nu_1 + \nu_2}{2}t' |it')}{\eta^6 (it')\,\theta_1 (- i
\nu_1 t'|it') \theta_1 (- i \nu_2 t'|it')},\nn
 &=& - \frac{2^4 \,n_1 n_2
\, V_{p + 1} \, \tan \pi \nu_1 \,\tan \pi\nu_2}{ (8 \pi^2
\alpha')^{\frac{1 + p}{2}}} \int_0^\infty \frac{d t'}{t'} \, e^{-
\frac{Y^2 t'}{2 \pi \alpha' }} \, t'^{  \frac{3 -
p}{2}}\frac{\sin^2(- i\pi \frac{\nu_1 - \nu_2}{2} t') \sin^2(- i \pi
\frac{\nu_1 + \nu_2}{2} t')}{\sin(- i\pi \nu_1 t') \sin (- i \pi
\nu_2 t')} \nn &\,&\qquad \times \prod_{n = 1}^\infty \frac{1}{(1 -
|z|^{2n})^4} \prod_{j = 1}^2 \frac{(1 - e^{\pi (\nu_1 + (-)^j
\nu_2)t'} |z|^{2n})^2 (1 - e^{- \pi  (\nu_1 + (-)^j \nu_2)t'}
|z|^{2n})^2}{(1 - e^{2\pi  \nu_j t'} |z|^{2n})(1 - e^{- 2\pi \nu_j
t'} |z|^{2n})},\eea where $|z| = e^{-\pi t'}$. We once again follow
\cite{Green:1996um, Lu:2009yx} to discuss the analytic structure of
the above amplitude and the associated physics. For the present (m,
m) case, both $\nu_1$ and $\nu_2$ are real with their respective
range $0 < \nu_1, \nu_2 < 1/2$ and the above amplitude appears
positive and has no simple poles on the positive $t'$-axis. For the
same reason mentioned in the previous subsection, this amplitude has
no imaginary part, therefore giving zero rate of open string pair
production as expected. Note that this amplitude has also a
singularity as $t' \rightarrow \infty$ when $Y \leq \pi \sqrt{2
|\nu_1 - \nu_2| \alpha'}$, i.e., on the order of string scale, and
this happens also in a similar fashion as in the brane/antibrane
system mentioned in previous subsection but now caused purely by the
presence of magnetic fluxes. This singularity can be examined from
\bea \lim_{t' \rightarrow \infty} \frac{ e^{- \frac{Y^2 t'}{2\pi
 \alpha'}}
\theta^2_1(\frac{ \nu_1 - \nu_2}{2 i} t'|it')\, \theta^2_1 (
\frac{\nu_1 + \nu_2}{2 i}t' |it')}{- \eta^6 (it')\,\theta_1 (-i
\nu_1 t'|it') \theta_1 (- i \nu_2 t'|it')} &\sim& \lim_{t'
\rightarrow \infty} \frac{e^{- \frac{Y^2 t'}{2\pi \alpha'}}
\sin^2(\pi \frac{\nu_1 - \nu_2}{2 i} t') \sin^2(\pi \frac{\nu_1 +
\nu_2}{2 i} t')}{ i^2 \sin(- i\pi \nu_1 t') \sin (- i \pi \nu_2
t')},\nn  &\sim& \lim_{t'\rightarrow\infty} e^{- \frac{Y^2 t'}{2\pi
\alpha'}}[e^{\pi |\nu_1 - \nu_2| t'}  + {\cal O} (1)]. \eea The
appearance of the divergent amplitude also indicates the breakdown
of the calculations, signalling the  onset of tachyonic instability
caused by the magnetic fluxes and the relaxation of the system to
form a new non-threshold bound state. In addition, that the open
string tachyon mode appears to arise is also indicated from the
leading term $e^{\pi |\nu_1 - \nu_2| t'}$, which diverges in the
short cylinder limit $t' \rightarrow \infty$, in the expansion of
the $\theta$-functions and $\eta$-function in (76) in the open
string channel. This is supported further by the evidence that this
divergence, therefore the tachyon mode, disappears when $|\nu_1 -
\nu_2|$  vanishes but when this happens the amplitude also vanishes,
indicating the underlying system being BPS and preserving certain
number of spacetime supersymmetries as mentioned earlier. Once
again, however, the detail of the underlying dynamical process
requires further understanding.

For the remaining case,  $\nu_1 = i \nu_{10}$ is imaginary with $0 <
\nu_{01} < \infty$ and $\nu_2 = \nu_{20}$ is real with $ 0 <
\nu_{20} < 1/2$. Then from the second expression in (76), we have
the amplitude \bea \Gamma &=&  \frac{ 4 \,n_1 n_2 \, V_{p + 1} \,
\tanh \pi \nu_{10} \,\tan \pi\nu_{20}}{ (8 \pi^2 \alpha')^{\frac{1 +
p}{2}}} \int_0^\infty \frac{d t'}{t'} \, e^{- \frac{Y^2 t'}{2 \pi
\alpha' }} \, t'^{ \frac{3 - p}{2}}\frac{(\cos\pi\nu_{10} t' - \cosh
\pi\nu_{20} t')^2}{\sin(\pi \nu_{10} t') \sinh ( \pi \nu_{20} t')}
\nn &\,&\times \prod_{n = 1}^\infty \frac{\prod_{j = 1}^2 (1 -
e^{\pi (i\nu_{10} + (-)^j \nu_{20})t'} |z|^{2n})^2 (1 - e^{- \pi
(i\nu_{10} + (-)^j \nu_{20})t'} |z|^{2n})^2}{(1 - |z|^{2n})^4 (1 - 2
|z|^{2n} \cos2\pi\nu_{10} t'
 + |z|^{4n})(1 - e^{2\pi \nu_{20} t'} |z|^{2n})(1 - e^{-
2\pi \nu_{20} t'} |z|^{2n})},\nn\eea where the following factor is
positive and can be expressed as \bea &&\prod_{j = 1}^2 (1 - e^{\pi
(i\nu_{10} + (-)^j \nu_{20})t'} |z|^{2n})^2 (1 - e^{- \pi (i\nu_{10}
+ (-)^j \nu_{20})t'} |z|^{2n})^2\nn && = [(1 + |z|^{4n})(1 +
|z|^{4n} - 4 |z|^{2n} \cos \pi \nu_{10} t \cosh\pi\nu_{20} t') + 2
|z|^{4n} (\cos 2\pi\nu_{10} t' + \cosh2\pi\nu_{20} t')]^2.\nn \eea
When $\nu_{20} \neq 0$ as we always assume in this subsection, the
above amplitude has simple poles occurring at $t'_k = k /\nu_{10}$
with $k = 1, 2, \cdots$ and the number of simple poles in the
present case doubles in comparison with the case when the two fluxes
share at least one common direction as discussed in the previous
subsection and in \cite{Bachas:1992bh, Bachas:1992zr, Bachas:1995kx,
Green:1996um, Cho:2005aj, Chen:2008jx, Lu:2009yx}\footnote{In some
of these papers, the number of simple poles appeared also given by
$t'_k = k/\nu_{01}$ with $k = 1, 2, \cdots$ due to that their
amplitude expressions were not simplified using the identity (42)
for various $\theta$-functions and the contribution to the amplitude
from each even $k$ is actually zero. This can also be seen easily
from (78) when taking $\nu_{20} = 0$.}. Then the rate of pair
production of open strings per unit worldvolume is the imaginary
part of the above amplitude, which is the sum of the residues of the
poles times $\pi$ following the prescription given in the previous
subsection as described in \cite{Bachas:1992bh, Bachas:1995kx}. We
then have the rate as \bea {\cal W} &\equiv& - \frac{2 {\rm Im}
\Gamma}{V_{p + 1}},\nn &=& \frac{8 n_1 n_2 \tanh \pi\nu_{10} \tan
\pi \nu_{20}}{\nu_{10}} \sum_{k = 1}^\infty (- )^{k + 1}
\left(\frac{\nu_{10}}{8 k \pi^2 \alpha'}\right)^{\frac{1 + p}{2}}
e^{- \frac{k Y^2}{2\pi \nu_{10} \alpha'}} \frac{\left[\cosh \frac{k
\pi \nu_{20}}{\nu_{10}} - (-)^k\right]^2}{\frac{\nu_{10}}{k} \sinh
\frac{k\pi \nu_{20}}{\nu_{10}}} \nn &\,&\qquad\times \prod_{n =
1}^\infty \frac{\left[1 - 2 (-)^k e^{- \frac{2 n k \pi}{\nu_{10}}}
\cosh \frac{k\pi \nu_{20}}{\nu_{10}} + e^{- \frac{4 n k
\pi}{\nu_{10}}}\right]^4}{\left[1 - e^{- \frac{2 n k
\pi}{\nu_{10}}}\right]^6 \left[1 - e^{- \frac{2 k \pi}{\nu_{10}}(n -
\nu_{20})}\right]\left[1 - e^{- \frac{2 k \pi}{\nu_{10}}(n +
\nu_{20})}\right]},\eea which reduces to the rate (55) given in the
previous subsection when we set $\nu_{20} \rightarrow 0$ and
$\nu_{10} = \nu_0$ in the above as expected. The rate is highly
suppressed by the separation and the integer $k$ and for each given
$k$ the corresponding term appears likely enhanced by both
$\nu_{10}$ and $\nu_{20}$. The latter is particularly evident for
large magnetic flux for which $\nu_{20} \rightarrow 1/2$ and the
front factor $\tan\pi \nu_{20} \rightarrow \infty$. Note that the
odd $k$ gives positive contribution while the even $k$ gives
negative contribution to the above rate. Also $k = 1$ term gives the
leading positive contribution to the rate\footnote{A different
enhancement of a similar rate by a magnetic flux in a different
context, i.e., purely bosonic string case, was explicitly considered
in \cite{Acatrinei:2000qm}. In this case, what has been considered
is an open string placed in an electric-magnetic background and the
two ends of the string obey only Neumann boundary conditions and
experience the same flux which can have both electric and magnetic
components, a generalization of the bosonic case considered in
\cite{Bachas:1992bh}.  The superstring case was also briefly
mentioned in \cite{Acatrinei:2000qm} but an explicit and compact
rate as ours was not given, neither was this rate enhancement
discussed. Further, the superstring rate there is for a special
case, i.e., for the space-filling $p = 9$ brane carrying an
electric-magnetic flux, which is excluded from our consideration. We
consider here a system of two sets of D$_p$ branes with a
separation, i.e. $p \le 8$, a more general and different system. In
our case, one end of the open string experiences an electric flux
living on one stack of D-branes while its other end experiences a
magnetic flux living on the other stack of D-branes placed parallel
at a separation, a superstring analysis. As a result, unlike the
case in \cite{Acatrinei:2000qm}, the present rate has a dependence
on the brane separation. Moreover, the enhancement factor explicitly
discussed in \cite{Acatrinei:2000qm} is merely a Born-Infeld factor
$\sqrt{1 + f_2^2}$ (expressed in our notations), independent of the
electric component, while the present enhancement is given as
$e^{\frac{\pi \nu_{20}}{\nu_{10}}} |f_2| /(8 \nu_{10})$ for fixed
$f_2$ with $\tan\pi \nu_{20} = |f_2|$, completely different. The
ratio of these two is $e^{\frac{\pi \nu_{20}}{\nu_{10}}} |f_2| /(8
\nu_{10} \sqrt{1 + f_2^2}) \approx e^{\frac{\pi \nu_{20}}{\nu_{10}}}
 /(8 \nu_{10}) \gg 1$ for $|f_2| \geq 1$ and small $\nu_{10}$. Even
 for the $p = 9$ case, our numerical estimation in the text shows that
 the corresponding rate should be negligibly small for a
 realistic  small electric flux in the presence of a fixed magnetic flux.} .
  All these indicate that the presence of magnetic
flux appears to enhance the rate. Let us consider the small
realistic electric flux case (Note that the flux is measured in
string units and so one expects a realistic flux to be small). For
this, we need to consider only the leading term which is given by
the $k = 1$ term in the above and for a fixed non-vanishing
$\nu_{20}$ it is\be {\cal W} \approx \frac{4 n_1 n_2 \pi}{\nu_{10}}
\left(\frac{\nu_{10}}{8\pi^2 \alpha'}\right)^{\frac{1 + p}{2}} e^{ -
\frac{Y^2}{2\pi \nu_{10} \alpha'}} e^{\frac{\pi \nu_{20}}{\nu_{10}}}
\tan \pi \nu_{20},\ee which is greatly enhanced by a factor of
$e^{\frac{\pi \nu_{20}}{\nu_{10}}} \tan \pi \nu_{20}/(8 \nu_{10})$
in comparison with the similar rate given in the previous
subsection. In particular, when the separation is on the order of
$\pi\sqrt{2\nu_{20}\alpha'}$, i.e., the string scale, this rate can
become very significantly large. In other words, when the two bound
states are in a close contact, the open string pair production can
be very significant. When this happens, we need to use the following
better approximated rate to make the evaluation \be {\cal W} =
\frac{4 n_1 n_2\, \pi\,\tan \pi \nu_{20}}{(8 \pi^2 \alpha')^{\frac{p
+ 1}{2}}} \sum_{k = 1}^\infty (- )^{k + 1} \left(\frac{\nu_{10}}{k
}\right)^{\frac{p - 1}{2}} e^{- \frac{k }{2\pi \nu_{10}
\alpha'}\left(Y^2 - 2 \pi^2 \nu_{20}\alpha'\right)}.\ee Let us make
some numerical estimation of the rate given in (81) when the
approximation is valid and this may serve for sensing the
significance of the rate mentioned above. For this purpose, we take
$\nu_{20} = 2/5, \,\nu_{10} = 1/50$ and the enhance factor given
above is then \be e^{\frac{\pi \nu_{20}}{\nu_{10}}} \,\frac{\tan \pi
\nu_{20}}{8 \,\nu_{10}} = e^{20 \pi} \frac{25 \tan 0.4 \pi}{4} \sim
3.6 \times 10^{28}.\ee We also calculate the rate in string units
for a few sample cases (note that we need to have $p \ge 3$ as
mentioned earlier) in the following as \bea (2 \pi \alpha')^{\frac{p
+ 1}{2}} {\cal W} &\approx& n_1 n_2 \left(\frac{\nu_{10}}{4
\pi}\right)^{\frac{p - 1}{2}} e^{ - \frac{Y^2 - 2\pi^2 \nu_{20}
\alpha'}{2\pi \nu_{10} \alpha'}} \tan \pi \nu_{20}\nn & \approx &
n_1 n_2 \left(\frac{\nu_{10}}{4 \pi}\right)^{\frac{p - 1}{2}}
\tan\pi\nu_{20},\nn
&\approx& \left\{\begin{array} {cc} 0.49  & {\rm for}\, p = 3, \\
0.03  & {\rm for}\, p = 4, \\
\end{array}\right.
\eea where we have taken $Y = \pi \sqrt{2\nu_{20} \alpha'} + 0^+
\approx 2.81 \sqrt{\alpha'}$,  $n_1 = 10$ and $n_2 = 10$. So the
rate can indeed be significant for $p = 3, 4$ and can be larger if
we take larger $n_1$ and $n_2$ when the brane separation is a few
times of string scale and before the tachyon condensation starts to
function. The above enhancement may have potentially realistic
applications, for example, to objects carrying both electric and
magnetic fluxes or to one object carrying an electric flux and the
other carrying a magnetic flux in a similar situation in our
Universe at early times or to the present macroscopic objects
carrying similar fluxes for which $n_1$ and $n_2$ are very large, if
string theories are indeed relevant to our real world. If this
indeed happens, the produced large number of open string pairs can
in turn annihilate to give highly concentrated high energy photons,
for example, which may have observational consequence such as the
Gamma-ray burst. The related effects may also serve as an indication
for the existence of extra dimensions since it requires $p \ge 3$
and the transverse dimensions are also necessary. Moreover, the rate
for $p = 3$ is at least one-order of magnitude larger than the other
$p \ge 4$, the underlying dynamics may select 4 spacetime dimensions
as special against the others, if a brane-world view is taken.  This
enhancement will not occur if the electric flux points along either
of the two spatial directions of the magnetic flux as our results
given in the previous subsection show. Note that  for small but
fixed electric flux as given in (81) or even  for a finite electric
flux as in (80),  the corresponding rate diverges when the magnetic
flux becomes large and this may indicate a new instability to occur.

There is another singularity which can be examined by looking at the
integrand of (78) at large $t'$ when $Y - \pi\sqrt{2\nu_{20}\alpha'}
\rightarrow 0^-$ as \be \lim_{t' \rightarrow \infty} \frac{ e^{-
\frac{Y^2 t'}{2\pi\alpha'}} (\cos\pi\nu_{10} t' - \cosh \pi\nu_{20}
t')^2}{ \sinh ( \pi \nu_{20} t')} \sim \lim_{t' \rightarrow \infty}
e^{- \frac{t'}{2\pi\alpha'} (Y^2 - 2 \pi^2 \nu_{20} \alpha')},\ee
which signals also the onset of tachyonic instability as in the pure
magnetic case (Note that this does not require a weak electric flux
and is associated with the real part of the amplitude).  For strong
electric flux, each term in (80) always diverges. So when $y
> \pi \sqrt{2\nu_{20} \alpha'}$, the pair production of open strings
is the only process to lower the system energy but as $Y \rightarrow
\pi \sqrt{2\nu_{20} \alpha'}$, the tachyonic instability starts to
occur and the pair production continues and and become larger and
larger. So the dynamics here may be rich and needs further study
before we can be certain to which the final state of this system
leads.

\section{Summary}
In this paper, we exhaust the amplitude calculations of
\cite{Lu:2009yx} between two non-threshold bound states of the type
of either ($F, D_p$) or ($D_{(p - 2)}, D_p$) or both for the
remaining cases as specified in the Introduction. We find that the
amplitude has the same basic structure when the two fluxes share at
least one common index. The amplitude is more general and  includes
the previous one as a special case when the two fluxes share no
common index. The nature of the force acting between  two bound
states is always attractive when the two fluxes are both magnetic or
magnetic dominant in a sense defined in the text.  For the rest of
cases considered in this paper, we are certain that the interaction
is attractive only at large separation. We also find that only for
two situations the interaction amplitude can vanish and if this
happens, the underlying system preserves 1/4 of spacetime
supersymmetries. One is that the two fluxes have different nature
with the magnetic flux sharing one common index with the electric
one, related to each other by (40), and the other is when the two
fluxes are both magnetic sharing no common index and having the same
magnitude.

We also study the analytic structures of various amplitudes
considered in this paper. When the two fluxes are both magnetic or
when the magnetic flux dominates over the electric flux in effect
and shares one common index, the amplitude diverges when the brane
separation is on the order of string scale just like the
brane/antibrane situation studied in \cite{Banks:1995ch, Lu:2007kv},
signalling the onset of tachyonic instability, and this may serve as
the dynamical process to lower the energy of the system and to relax
it to form the final stable bound state as indicated in
\cite{polbooktwo}.

For the rest of cases studied, i.e., the cases with at least one or
one dominant electric flux present,  there is always a non-vanishing
open string pair production rate associated with this flux. In
particular, this rate can be greatly enhanced when there is in
addition a magnetic flux present and the electric flux is weak but
with the two orthogonal to each other. These may have realistic
physical consequences for and potential applications to objects in
our Universe and their evolution when they carry a weak electric
field and a reasonable but fixed magnetic field and if string
theories are relevant to our real world. If this happens indeed, it
may serve also as an indication for the existence of extra
dimensions since the spatial dimensionality of the D$_p$ branes has
now to be $p \ge 3$. Further, our usual 4 dimensional spacetime
seemly has also a special role since the rate in this case is at
least one-order of magnitude larger than the other relevant cases.
Pursuing all these applications is beyond the scope of the present
work and will be postponed to future projects.

In addition to the usual strong electric flux singularity for the
pair production rate, for the present case we also find  two new
singularities when the brane separation is on the order of string
scale: one is associated with the pair production rate for a weak
electric flux with a large magnetic flux and the other is from the
real part of the amplitude and associated with the onset of
tachyonic instability but independent of the electric flux
requirement. The dynamics here may be rich and needs further study
before we can be sure about the final state of the system.

\vspace{.5cm}

\noindent {\bf Acknowledgements}
 \vspace{2pt}

We would like to thank Zhao-Long Wang and Rong-Jun Wu for useful
discussions. JXLU would like to thank the Kavli Institute for
Theoretical Physics at Beijing for hospitality during the final
stage of this work. We acknowledge support by grants from the
Chinese Academy of Sciences, a grant from 973 Program with grant No:
2007CB815401 and grants from the NSF of China with Grant No:10588503
and 10535060.

\appendix
\newcounter{pla}
\renewcommand{\thesection}{\Alph{pla}}
\renewcommand{\theequation}{\Alph{pla}.\arabic{equation}}
\setcounter{pla}{1} \setcounter{equation}{0}

\section*{Appendix}
 In this appendix, we will confirm explicitly the
preservation of 1/4 spacetime supersymmetries for each of the three
cases mentioned in the text when the corresponding amplitude
vanishes, namely, (e, m) when the electric flux shares a common
spatial direction with the magnetic flux along with (40) satisfied,
and the (m, m) case when the two magnetic fluxes share no common
direction along with .

For this, let us first note that the condition for 1/2-supersymmetry
preservation for each bound state is \be \epsilon_1 = \eta \,
\Gamma^{p + 1} \cdots \Gamma^9 U (\hat F_j) \,\epsilon_2,\ee where
$\epsilon_1$ and $\epsilon_2$ are the two Majorana-Weyl
supersymmetry parameters in IIA/IIB string theory (the two have the
opposite chirality in IIA but the same chirality in IIB), $U(\hat
F_j)$ is defined in (12) due to the presence of flux, the sign $\eta
= \pm$ and $j = 1, 2$. This condition reduces to the familiar one
when the flux is set to zero (note that we have used $\Gamma^0
\Gamma^1 \cdots \Gamma^p = \pm \,\Gamma^{p + 1} \cdots \Gamma^9
\Gamma^{11}$ and $\Gamma^{11} \epsilon_2 = \pm\, \epsilon_2$). The
above clearly indicates that the SUSY parameter $\epsilon_1$ is
completely determined by $\epsilon_2$, therefore only half
supersymmetry is preserved for a given bound state when it is
isolated, the well-known fact. For the cases under consideration, we
have two bound states with the Dp branes in one bound state placed
parallel to those in the other bound state at a separation.
Therefore, for such a system to preserve certain number of
supersymmetries, we need to have (A.1) hold simultaneously for $j
=1, 2$. This is equivalent to  having (A.1) hold for either $j = 1$
or $j = 2$  plus the following \be U(\hat F_1)\,\epsilon_2 = U(\hat
F_2))\,\epsilon_2\ee for non-vanishing $\epsilon_2$. The number of
non-vanishing components of $\epsilon_2$ satisfying the above
equation determines the number of unbroken SUSY for such a system.

 Let us check the (e, m) case first. For this case, \be
U (\hat F_1) = \frac{1 + f_1 \Gamma^0 \Gamma^1}{\sqrt{1 -
f_1^2}},\ee where we choose the electric flux along $x^1$-direction.
Without loss of generality we can choose the magnetic flux along
$x^1$ and $x^2$ directions and then \be U(\hat F_2) = \frac{1 + f_2
\Gamma^1 \Gamma^2}{\sqrt{1 + f_2^2}}.\ee So (A.2) now becomes  \be
\left(1 \pm {\cal B}\right)\, \epsilon_2 = 0, \ee where we have
expressed $f_2$ in terms of $f_1$ using  (40) and \be {\cal B} =
\sqrt{1 - f_1^2} \Gamma^0 \Gamma^2 \pm f_1 \Gamma^0 \Gamma^1.\ee
Since ${\rm Tr}\, {\cal B} = 0$ and ${\cal B}^2 = I_{32\times 32}$
with $I_{32\times 32}$ the unit matrix, (A.5) says that only half of
the components of $\epsilon_2$ can be non-vanishing and therefore
overall only 1/4 of the spacetime supersymmetries can be preserved
by this configuration. The (m, e) case can be similarly discussed
and the conclusion remains the same, i.e., the underlying system
preserves only 1/4 of overall spacetime supersymmetries.

We now move to the (m, m) case mentioned above. For this case, we
choose the first magnetic flux $\hat F_1$ along $x^1$ and $x^2$
directions and so \be U (\hat F_1) = \frac{1 + f_1 \Gamma^1
\Gamma^2}{\sqrt{1 + f_1^2}},\ee and without loss of generality we
choose the second magnetic flux $\hat F_2$ along $x^3$ and $x^4$
directions and so \be U (\hat F_2) = \frac{1 + f_2 \Gamma^3
\Gamma^4}{\sqrt{1 + f_2^2}}.\ee (A.2) now reduces to \be (1 \pm
\Gamma^1 \Gamma^2 \Gamma^3 \Gamma^4) \, \epsilon_2 = 0,\ee where we
have used $f_1 = \pm f_2$ which is precisely the one giving the
vanishing amplitude in subsection 3.2. Since ${\rm Tr}\, \Gamma^1
\Gamma^2 \Gamma^3 \Gamma^4 = 0$ and $(\Gamma^1 \Gamma^2 \Gamma^3
\Gamma^4)^2 = I_{32\times 32}$, by the same token, this system
preserves also 1/4 of total spacetime supersymmetries.

\end{document}